\documentclass[a4paper,11pt]{article}

\usepackage{jheppub} 
\usepackage{lineno}

\usepackage{graphicx}
\usepackage{bm}
\usepackage{mathtools}
\usepackage[table,dvipsnames]{xcolor}
\usepackage[compat=1.1.0]{tikz-feynman} 
\usepackage{float} 
\usepackage{xfrac}
\usepackage{subcaption}

\title{\boldmath Minimal extension of the Standard Model\\ with a mirror symmetry\\ between fundamental fermions\\ and a possible origin of dark matter 
}

\author{Jakub Rembieliński}
\affiliation{University of Lodz, Faculty of Physics and Applied Informatics,\\Pomorska 149/153, 90-236 Lodz, Poland}

\emailAdd{jaremb@uni.lodz.pl}

\abstract{
In this paper, we propose a specific, nontrivial extension of the Standard Model of weak interactions based on the ${SU(2)}_L\times{U(1)}_Y\times{U(1)}_C$ group. Our motivation follows from the identification of the globally conserved charge  $\Omega=\mathrm{B}-\mathrm{L}-\mathrm{Q}$ as a neutrino charge. Here $\mathrm{Q}$ denotes the electric charge, $\mathrm{L}$ is the lepton number, and $\mathrm{B}$ is the baryon number. We introduce a corresponding gauge field, $\mathit{\Omega}_\mu$, that interacts solely with the neutrinos in the lepton sector. Next, we extended this scheme to the quark sector, constructing the corresponding Lagrangians and currents. An intriguing feature of the model is the emergence of mirror symmetry between neutrinos and electrically charged leptons, as well as between up and down quarks. Following the spontaneous breaking of the weak gauge symmetry with the use of the Goldstone-Higgs iso-doublet, all elementary fermions acquire Dirac masses from Yukawa interaction. The $W^\pm$ and $Z$ bosons also acquire masses, although with a modified relationship between their respective masses, as compared to the Standard Model. Our model accounts for both chiral components of the neutrino and offers an explanation for the non-observability of the right-chiral neutrino. Additionally, it forbids neutrinoless double-beta decay. Spontaneous breaking of the local $U(1)_\Omega$ symmetry leads to the gauge boson mass $M_\mathit{\Omega}$, which we assume to be greater than the mass $m_\chi$  of a new scalar, Higgs-like field $\chi$. The cosmological stability of $\chi$, predicted under this condition, allows for its interpretation as dark matter, interacting exclusively with $\mathit{\Omega}_\mu$ and gravity. The $\mathit{\Omega}_\mu$ particles also interact with Standard Model particles. Besides the Standard Model parameters and the masses $m_\chi$ and $M_\mathit{\Omega}$, our framework introduces a new, feeble coupling $q$, with $q\ll g$ (where $g$ is the electroweak interaction coupling). From this perspective, we solve and analyze a system of Boltzmann equations that describe the thermal evolution of the number density of $\chi$  dark matter and the $\mathit{\Omega}_\mu$ mediator field within the context of the $\mathrm{\Lambda CDM}$ cosmological model. By employing the freeze-in mechanism, we identify regions in the parameter space of the model that yield the observed relic abundance of dark matter. Specifically, we estimate the coupling constant $q$ to be in the range of  $\sim 10^{-8.5}\ g$ to $\sim 10^{-6}\ g$ which ensures cosmological stability of the $\chi$ particles.
}

\makeatletter 
\gdef\@fpheader{}
\makeatother

\begin{document}
\maketitle
\flushbottom

\section{Introduction} \label{Sec:intro}

The Standard Model (SM) is one of the most successful achievements in modern physics, offering highly precise descriptions of particle interactions, particularly in the electroweak sector. Despite its success, several important questions remain unresolved, especially concerning the nature of neutrinos. These include the origin of their mass, the mechanism for the decoupling of right-chiral neutrinos, and the breaking of CP symmetry. They are still waiting for conclusive answers. To address these questions, several generalizations of the Standard Model have been proposed often through extensions of the electroweak group. In recent decades many $U(1)$ extensions of the SM have been studied typically involving the spontaneous breaking of $U(1)$ symmetry by means of an additional scalar Higgs field and with one additional massive vector boson $Z'$ \cite{Applequist2003Nonexotic,Langacker2009ThePhysics,Ekstedt2018Minimal,Arcadi2018TheWaning,Kors2004AStuckelberg,Burgess2001TheMinimal}. A class of these extensions has focused on the dark sector, often involving a dark photon \cite{Dobrescu2005Masslss,Dobrescu2006Spin,Fabbrichesi2020TheDark,An2015Direct,Kahn2017Light,Dienes2012Dynamical,Heeck2014Unbroken}. 

In this paper, we propose a specific minimal $ U(1) $ extension of the original electroweak model, preserving its fundamental structure and pattern. Moreover, we investigate some of its cosmological consequences, like dark matter genesis and its abundance. Our motivation follows from the identification of the globally conserved charge
\begin{equation} \label{eq:Intro1}
    \Omega = \mathrm{B}-\mathrm{L}-\mathrm{Q},
\end{equation}
as a neutrino charge. Here $\mathrm{Q}$ is the electric charge, $\mathrm{L}$ is the lepton number and $\mathrm{B}$ is the baryon number. According to \eqref{eq:Intro1}, neutrinos possess a charge $ \Omega = -1 $ while for electrically charged leptons $ \Omega = 0 $. In the quark generations case, for up quarks $ \Omega = -\frac{1}{3} $ while for down quarks $ \Omega = +\frac{2}{3} $. Notice that this $\mathrm{\Omega}$ charge pattern is a mirror image of the electric one. However, in the Standard Model, in contrast to the electric charge, this globally conserved charge is not related to a corresponding local current. For this reason, our crucial assumption is that in the lepton sector a gauge field ($\mathit{\Omega}_\mu$) of the investigated SM extension is coupled solely to the neutrinos. We also identify the lepton and quark currents coupled to $\mathit{\Omega}_\mu$. As a result, we obtain a surprising mirror symmetry between neutrinos and charged leptons as well as between up and down quarks. Our model incorporates both left- and right-chiral neutrinos with Dirac masses and explains the experimental non-observability of their right component. The model (extended to three generations) is free of chiral and gravitational gauge anomalies, so it is renormalizable. It is worth stressing that the model is constructed without prior assumption of gauge anomalies cancellation.

The interaction predicted by the model produces a repulsive force on baryonic matter containing neutrons because for neutrons $ (udd) $, $ \Omega = +1 $ while for protons $ (uud) $, $ \Omega = 0 $. To avoid a conflict with gravity in the case of the long-range field (massless $\mathit{\Omega}_\mu$), this force should be extremely small, or the $\mathit{\Omega}_\mu$ field should becomes massive. In this paper, we consider the latter possibility. Therefore, we use the standard spontaneous symmetry breaking mechanism to generate a nonzero mass for the  $\mathit{\Omega}_\mu$. As a result,  $\mathit{\Omega}_\mu$ vector boson couples to an additional, dark Higgs-like, iso-singlet, scalar field $\chi$, which is stable in the cosmological scale and interacts solely with $\mathit{\Omega}_\mu$ allowing it to be interpreted as a dark matter field. The stability of dark particles is ensured by an inverted, spontaneously generated mass hierarchy in the dark sector and estimated weakness of the new force. The gauge field $\mathit{\Omega}_\mu$ serves as a mediator field between the dark sector and the SM particles. We explore the generation of the dark matter abundance within the context of the freeze-in scenario of the $\mathrm{\Lambda CDM}$ cosmological standard model. This requires an appropriate determination of the interaction strengths between $\mathit{\Omega}_\mu$ and $\chi$ interactions. For simplicity, we limit our discussion to one generation of leptons denoted  as $(\nu,l)$ representing $\left(\nu_l,l\right)$, where $l=e,\mu,\tau$, and one generation of quarks $(u,d)$, where $u=u,c,t$ and $d=d,s,b$. A generalisation to the three-generations case is straightforward.

\section{Preliminaries} \label{Sec:prelim}

The electroweak group of the Standard Model is identified with the direct product $ SU(2)_{L}\times U(1)_{Y} $ under a number of physically well-founded assumptions. Here $Y$ is the weak hypercharge while $T$ denote the weak isospin. As we know, the group $SU(2)_{L}\times U(1)_{Y}$ is differently realized on the left$(L)$- and right$(R)$-chiral doublets: on $ L $ as faithful representation (iso-doublet) while on $R$ as a pair of one-dimensional realizations of $U(1)_{Y}$ only. In the Weyl bi-spinor representation the lepton and quark fields split on the chiral left and right components

\begin{subequations} \label{eq:Prelim1}
    \begin{align} 
    \label{subeqn:Prelim1a}
        \begin{matrix}
            \mathrm{L} &=\begin{pmatrix} \nu_{L} \\ l_{L} \\ \end{pmatrix} 
        & 
            \textrm{or} 
        & 
            \begin{pmatrix} u_{L} \\ d_{L} \\ \end{pmatrix} 
        \end{matrix},        
    \\
    \label{subeqn:Prelim1b}
        \begin{matrix}
            \mathrm{R} &=\begin{pmatrix} \nu_{R} \\ l_{R} \\ \end{pmatrix} 
        & 
            \textrm{or} 
        & 
            \begin{pmatrix} u_{R} \\ d_{R} \\ \end{pmatrix} 
        \end{matrix}.
    \end{align}
\end{subequations}

In this paper we examine an extension of $ SU(2)_{L}\times U(1)_{Y} $ using the fact that in the space of the chiral fields we can nontrivially implement an additional $ U(1) $ group i.e. we extend the electroweak group to $ SU(2)_{L}\times U(1)_{Y}\times U(1)_{C} $ with the following, most general, realization in the chiral space

\begin{subequations} \label{eq:Prelim2}
\begin{align} 
\label{subeqn:Prelim2a}
    \mathrm{L}' &= e^{\displaystyle i\frac{\boldsymbol{\alpha} \boldsymbol{\tau}}{2}} e^{\displaystyle i\frac{y_{1}\beta }{2}} e^{\displaystyle i\frac{c_{1}\gamma }{2}} \mathrm{L},
\\
\label{subeqn:Prelim2b}
    \mathrm{R}' &=\begin{pmatrix}
    e^{\displaystyle i\frac{y_{2}\beta }{2}} e^{\displaystyle i\frac{c_{2}\gamma }{2}} & 0 \\ 
    0 & e^{\displaystyle i\frac{y_{3}\beta }{2}} e^{\displaystyle i\frac{c_{3}\gamma }{2}} 
    \end{pmatrix} \mathrm{R},
\end{align}
\end{subequations}

\noindent where iso-vector $ \boldsymbol{\alpha} $ parametrize the weak isospin group while $ \beta $ parametrize $ U(1)_{Y} $ with corresponding hypercharge values $\left\{ y_{1},y_{2}, y_{3}\right\} $, different for leptons and quarks and $\boldsymbol{\tau} $ is the Pauli matrix triplet. Here $ \gamma $ parametrize the additional $ U(1)_{C} $ group while generators (C-charges) $\left\{ c_{1}, c_{2}, c_{3}\right\} $ fix one dimensional representations of this group for leptons and $ \left\{\tilde{c}_{1},\tilde{c}_{2},\tilde{c}_{3}\right\} $ for quarks. We will use the relationship between electric charge and weak hypercharge and isospin in the usual form
\begin{equation} \label{eq:Prelim3}
    Q = T_{3}+\frac{1}{2}Y. 
\end{equation}

\section{Gauging of the extended electroweak group} \label{Sec:gauging}

We apply the standard procedure of gauging of the extended electroweak group \break $ {SU(2)_{L}\times U(1)_{Y}\times U(1)_{C}} $. Except of the gauge fields $ \bm{W}_{\mu } $ related to $ SU\left(2\right)_{L} $ and $ B_{\mu } $ related to $ U\left(1\right)_{Y} $ we introduce an additional gauge field $ C_{\mu } $ related to $ U(1)_{C} $. The corresponding covariant derivatives take the form $ iD_{\mu }^{L} = i\partial_{\mu }+\Gamma_{\mu }^{L} $, $ iD_{\mu }^{R} = i\partial_{\mu }+\Gamma_{\mu }^{R} $ with

\begin{subequations} \label{eq:Gauging1}
    \begin{align}
        \label{subeqn:Gauging1a}
        \begin{split}
            \Gamma_{\mu }^{L} &=\frac{g}{\sqrt{2}}\begin{pmatrix} 0 & W_{\mu }^{+} \\ W_{\mu }^{-} & 0 \\ \end{pmatrix}+\frac{g}{2}\begin{pmatrix} W_{\mu }^{3} & 0 \\ 0 & -W_{\mu }^{3} \end{pmatrix}+
            \frac{y_{1}g'}{2}\begin{pmatrix} B_{\mu } & 0 \\ 0 & B_{\mu } \\ \end{pmatrix}+\frac{c_{1}g''}{2}\begin{pmatrix} C_{\mu } & 0 \\ 0 & C_{\mu } \end{pmatrix},
        \end{split}
    \\
        \label{subeqn:Gauging1b}
        \Gamma_{\mu }^{R} &=\frac{ g'}{2}\begin{pmatrix} y_{2}B_{\mu } & 0 \\ 0 & y_{3}B_{\mu } \\ \end{pmatrix}+\frac{g''}{2}\begin{pmatrix} c_{2}C_{\mu } & 0 \\ 0 & c_{3}C_{\mu } \\ \end{pmatrix},
    \end{align}
\end{subequations}

\noindent where for leptons $\left\{y_{1},y_{2}, y_{3}\right\}=\left\{-1,0,-2\right\}$ and for quarks $\left\{y_{1},y_{2}, y_{3}\right\}=\left\{\frac{1}{3},\frac{4}{3},-\frac{2}{3}\right\}$ while $c_{i }\rightarrow\tilde{c}_{i}$. The charged vector bosons are denoted as usually, i.e. $W^\pm=\left(W_1\mp i W_2\right)/\sqrt{2}$. The corresponding gauge coupling constants are denoted as $g$, $g'$, and $g''$.

\section{The lepton sector} \label{Sec:lepton}

Our main assumption is that in the lepton sector the physical gauge field $\mathit{\Omega}_\mu$ couples solely with neutrinos. We also use the obvious physical requirement that the electromagnetic field $A_\mu$ couples in the fermionic sectors with the charged leptons and quarks in the standard way. To determine the connection coefficients \eqref{eq:Gauging1} in our case, we begin with a general orthogonal relationship between the gauge fields $W_\mu^3,\ B_\mu, C_\mu$ and the physical fields $A_\mu,\ Z_\mu,\ \mathit{\Omega}_\mu$ and eliminate $W_\mu^3,\ B_\mu, C_\mu$ from connections $\Gamma_\mu^L$ and  $\Gamma_\mu^R$. Next, we multiply by the connection $\Gamma_\mu^L$ -- the lepton doublet $L$ and by $\Gamma_\mu^R$ -- the doublet $R$. As a result of elimination couplings of $A_\mu$ with neutrino field $\nu$ as well as couplings of $\mathit{\Omega}_\mu$ with charged lepton field $l$, we obtain four conditions on the coupling constants and the C-charges. Two additional conditions arise from the fact that left and right charged leptons $l_L$ and  $l_R$  have the same electric charge $\mathrm{Q}$ (so the same coupling with $A_\mu$) while left and right neutrinos  $\nu_L$  and  $\nu_R$  have the same neutrino charge $\Omega$ (so the same coupling with $\mathit{\Omega}_\mu)$.  Moreover, the electric coupling and values of the weak hypercharge of fundamental fermions should be the same as in the Standard Model. Taking the above into account, we obtain that the orthogonal relationship between gauge fields has the form

\begin{subequations} \label{eq:Lepton1}
    \begin{align}
        \label{subeqn:Lepton1a}
        W_{\mu }^{3} &= Z_{\mu }\cos \theta \cos \varphi +A_{\mu }\sin \theta -\mathit{\Omega}_\mu\cos \theta \sin \varphi,
    \\
        \label{subeqn:Lepton1b}
        B_{\mu } &= -Z_{\mu }\sin \theta \cos \varphi +A_{\mu }\cos \theta +\mathit{\Omega}_\mu\sin \theta \sin \varphi,
    \\
        \label{subeqn:Lepton1c}
        C_{\mu } &= Z_{\mu }\sin \varphi +\mathit{\Omega}_\mu\cos \varphi,
    \end{align}
\end{subequations}

\noindent as well as we find the following relations between the gauge couplings and $C$-charges: 
\begin{equation} \label{eq:Lepton2}
\begin{split}
    \dfrac{g'}{g} &= \tan \theta,
\\
    c_{1}\left(\frac{g''}{g}\right) &= -\cos \theta\left(1-\tan^{2}\theta\right)\tan \varphi,
\\
    c_{2}\left(\frac{g''}{g}\right) &= -2\cos \theta \tan \varphi,
\\
    c_{3}\left(\frac{g''}{g}\right) &= 2\sin \theta \tan \theta \tan\varphi,
\end{split}
\end{equation} 
so
\begin{equation} \label{eq:Lepton3}
    2c_{1} = c_{2}+c_{3}.
\end{equation}
Evidently, the angle $\theta$ is identified as the SM Weinberg angle while $\phi$ is a new mixing angle governed the new interaction strength. The explicit form of  C-charges will be determined in the Sec.\ref{Sec:mirror}. Now, considering the bi-spinor character of the neutrino and charged lepton fields and using  the Eqs. \eqref{eq:Gauging1}, \eqref{eq:Lepton1} and \eqref{eq:Lepton2}, the connection coefficients for neutrino and charged lepton, take the following form

\begin{subequations} \label{eq:Lepton4}
    \begin{align}
        \begin{split} \label{subeqn:Lepton4a}
            \Gamma_{\mu }^{\nu } &= -g\mathit{\Omega}_\mu\cos\theta \sin\varphi\,\mathrm{I}+\frac{g Z_{\mu }}{2\cos\theta\cos\varphi}\left(\frac{\left(1-4\cos^{2}\theta\sin^{2}\varphi\right)}{2}\mathrm{I}-\frac{1}{2}\gamma^{5}\right),
        \end{split}
        \\
        \begin{split} \label{subeqn:Lepton4b}
            \Gamma_{\mu }^{l} &= -gA_{\mu }\sin\theta\,\mathrm{I}+ 
            \frac{g Z_{\mu }}{2\cos\theta\cos\varphi}\Bigg(\left( \frac{\left(4\sin^{2}\theta-1\right)}{2}+\sin^{2}\varphi \tan^{2}\theta\left(2 \sin^{2}\theta-1\right)\right)\mathrm{I}+
            \\
            &+\left(\frac{1}{2}+\sin^{2}\varphi \tan^{2}\theta\left(2 \sin^{2}\theta-1\right)\right)\gamma^{5}\Bigg),
        \end{split}
        \\
        \begin{split} \label{subeqn:Lepton4c}
            \Gamma_{\mu }^{\mathrm{off}} &= \frac{g}{2\sqrt{2}}
            \begin{pmatrix}
                0 & W_\mu^+\left(\mathrm{I}-\gamma^5\right) \\
                W_\mu^-\left(\mathrm{I}-\gamma^5\right) & 0
            \end{pmatrix}
        \end{split}
    \end{align}
\end{subequations}

\noindent where the identity $\mathrm{I}$ and $\gamma^5$ are elements of the Clifford algebra generated by $\gamma^\mu$ matrices. The term $\Gamma^\mathrm{off}$ corresponds to the off-diagonal part of the connection \eqref{subeqn:Gauging1a}.

Finally, inserting in the free massless lepton field Lagrangian the calculated covariant derivatives, and taking into account the effect of mass generation from Yukawa coupling with the standard Higgs field $ H $, we obtain the lepton Lagrangian in the form

\begin{equation} \label{eq:Lepton5}
    \begin{split}
        \mathcal{L}_{\mathrm{lepton}} &=\overline{\nu }\gamma^{\mu }i\partial_{\mu }\nu -m_{\nu }\overline{\nu }\nu +\overline{l}\gamma^{\mu }i\partial_{\mu }l-m_{l}\overline{l}l
        -e A_{\mu }j_{l}^{\mu }-q\mathit{\Omega}_\mu  j_{\nu }^{\mu }
        -\frac{g}{2\sqrt{2}}\left(W_{\mu }^{+}j_{-}^{\mu }+W_{\mu }^{-}j_{+}^{\mu }\right)+
    \\
        &-\frac{g Z_{\mu }}{2\cos\theta\cos\varphi}\Big(\left(g_{V}^{\nu } j_{\nu }^{\mu }-g_{A}^{\nu }j_{5\nu }^{\mu }\right)
        +\left(g_{V}^{l}j_{l}^{\mu }-g_{A}^{l}j_{5l}^{\mu }\right)\Big)
        -\frac{m_{l}}{\vartheta }\overline{l}Hl-\frac{m_{\nu }}{\vartheta }\overline{\nu }H\nu,
    \end{split}
\end{equation}    

\noindent where $ e=g \sin \theta $, $ q = g \cos \theta \sin \varphi $ are the values of the corresponding coupling constants and $ \vartheta^{2} =1 / \sqrt{2}G_{F} $, $\vartheta $ is vacuum expectation value (VEV) of the Goldstone–Higgs iso-dublet. The currents in Eq.\eqref{eq:Lepton5} have the following form 

\begin{equation} \label{eq:Lepton6}
    \begin{aligned}
        j_{\nu }^{\mu } &=\overline{\nu } \gamma^{\mu }\nu, & \quad
        j_{5\nu }^{\mu } &=\overline{\nu }\gamma^{\mu } \gamma^{5} \nu, & \quad
        j_{+}^{\mu } &=\overline{l}\gamma^{\mu }\left(1-\gamma^{5}\right)\nu, \\
        j_{l}^{\mu } &=\overline{l} \gamma^{\mu } l, & \quad
        j_{5l}^{\mu } &=\overline{l} \gamma^{\mu } \gamma^{5} l, & \quad
        j_{-}^{\mu } &=\overline{\nu }\gamma^{\mu }\left(1-\gamma^{5}\right)l.
    \end{aligned}
\end{equation}

The coupling constants $ g_{V/A}^{\nu /l} $ in the Lagrangian \eqref{eq:Lepton5} are given by the formulas
\begin{equation} \label{eq:Lepton7}
    \begin{aligned}
        g_{V}^{\nu } &=\frac{\left(1-4\sin^{2}\varphi\cos^{2}\theta\right)}{2}, & \quad
        g_{A}^{\nu } &=\frac{1}{2}, \\
        g_{V}^{l} &=\frac{1}{2}+\left(2\sin^{2}\theta-1\right)\left(1+\sin^{2}\varphi \tan^{2}\theta\right), &  \quad      
        g_{A}^{l} &= -\left(\frac{1}{2}+\sin^{2}\varphi \tan^{2}\theta\left(2\sin^{2}\theta-1\right)\right).
    \end{aligned}
\end{equation}
 
For $ \varphi \rightarrow 0 $ the above formulas turn into standard SM form. As we see from Eq.\eqref{eq:Lepton5}, in the lepton sector $ \mathit{\Omega}_\mu  $ is coupled with the neutrino vector current $ j_{\nu }^{\mu } =\overline{\nu } \gamma^{\mu } \nu $ and the value of the coupling constant $q $ is given by equality $q = g \cos \theta \sin \varphi =e\cot \theta \sin \varphi $. As in the standard case, the right chiral neutrino does not interact with charged gauge bosons $ W^{\pm } $ because of projection on the left chiral neutrino in the interaction term. However, $ \nu_{R} $ together with $ \nu_{L} $ is involved in interaction with $ \mathit{\Omega}_\mu  $ and $ Z_{\mu } $ bosons with strength dependent on the value of the angle $ \varphi $ (see Eqs.\eqref{eq:Lepton5}, \eqref{eq:Lepton6} and \eqref{eq:Lepton7}). As we will see (Sec.\ref{Sec:evolution}), the coupling constant $q$ is very small so also $\varphi$, which implies that the constant $g_V^\nu\rightarrow\frac12$ so $\nu_R$ interaction with gauge bosons is negligible. This is the reason for the non-observability of the right-chiral neutrino.

We see from the Lagrangian \eqref{eq:Lepton5}, that the global charge $ \Omega $ takes value $-1$ $(+1)$ for the neutrino $ \nu $ $\left(\overline{\nu }\right)$ and $ 0 $ for the charged lepton $ l $ $\left(\overline{l}\right) $, in accordance with the definition \eqref{eq:Intro1}.

\section{The quark sector} \label{Sec:quark}

In the quark sector, we obtain the realization of the electroweak group $ SU(2)_{L}\times U(1)_{Y}\times U(1)_{C} $ in the following form

\begin{subequations} \label{eq:Quark1}
    \begin{align}
        \label{subeqn:Quark1a}
        \begin{pmatrix}
        u_{L}^{'} \\ 
        d_{L}^{'} \\ 
        \end{pmatrix} &= e^{\displaystyle i\frac{\boldsymbol{\alpha}\boldsymbol{\tau}}{2}}e^{\displaystyle i\frac{\beta }{6}}e^{\displaystyle i\frac{\tilde{c}_{1}\gamma }{2}}\begin{pmatrix}
        u_{L} \\ 
        d_{L} \\ 
        \end{pmatrix},
    \\
        \label{subeqn:Quark1b}
        \begin{pmatrix}
        u_{R}^{'} \\ 
        d_{R}^{'} \\ 
        \end{pmatrix} &=\begin{pmatrix}
        e^{\displaystyle i\frac{2\beta }{3}}e^{\displaystyle i\frac{\tilde{c}_{2}\gamma }{2}}u_{R} \\ 
        e^{\displaystyle -i\frac{\beta }{3}}e^{\displaystyle i\frac{\tilde{c}_{3}\gamma }{2}}d_{R} \\
        \end{pmatrix}.
    \end{align}
\end{subequations}

A procedure analogous to the lepton case provides the following relations between $c$-, $\tilde{c}$-charges and gauge couplings:
\begin{equation} \label{eq:Quark2}
\begin{split}
    \tilde{c}_{1}\left(\frac{g''}{g}\right) &=\frac{1}{3}\cos \theta\left(1-\tan^{2}\theta\right)\tan \varphi, \\
    \tilde{c}_{2}\left(\frac{g''}{g}\right) &= -\frac{2}{3}\cos \theta (1+2 \tan^{2}\theta )\tan \varphi, \\
    \tilde{c}_{3}\left(\frac{g''}{g}\right) &= \frac{2}{3}\cos \theta (2+  \tan^{2}\theta )\tan \varphi,
\end{split}
\end{equation}
leading to the relations between $\tilde{c}$ and $c$:
\begin{equation} \label{eq:Quark3}
\begin{aligned}
    \tilde{c}_{1} &= -\frac{1}{3}c_{1}, & \quad
    \tilde{c}_{2} &= c_{2}-\frac{4}{3}c_{1}, & \quad
    \tilde{c}_{3} &=\frac{2}{3} c_{1}-c_{2},
\end{aligned}
\end{equation}
so
\begin{equation} \label{eq:Quark4}
    2\tilde{c}_{1} =\tilde{c}_{2}+\tilde{c}_{3}.
\end{equation}

Thus by means of Eqs. \eqref{eq:Gauging1}, \eqref{eq:Lepton2} and \eqref{eq:Quark3}, the diagonal connection coefficients take the form

\begin{subequations} \label{eq:Quark5}
    \begin{align}
        \label{subeqn:Quark5a}
        \begin{split}
            \Gamma_{\mu }^{u} &=\frac{2}{3}g A_{\mu }\sin\theta\,\mathrm{I}+ \\
            &+\frac{g}{2 \cos\theta \cos\varphi}\frac{Z_\mu}{2} \left(\left(\left(1-\frac{8}{3} \sin^{2}\theta\right)\cos^{2}\varphi-\frac{4}{3}\sin^{2}\theta\sin^{2}\varphi\right)\mathrm{I}-\gamma^{5}\right)+ \\ 
            &-\frac{1}{3}g \mathit{\Omega}_\mu\cos\theta\sin\varphi\,\mathrm{I},
        \end{split}
        \\
        \label{subeqn:Quark5b}
        \begin{split}
            \Gamma_{\mu }^{d} &= -\frac{1}{3}g A_{\mu }\sin\theta\,\mathrm{I}+\frac{Z_{\mu }}{2}\left(\left(\left(-1 +\frac{4}{3} \sin^{2}\theta\right)\cos^{2}\varphi+\frac{1}{3}\sin^{2}\varphi\right)\mathrm{I}+\gamma^{5}\right)+ \\
            &+\frac{2}{3}g \mathit{\Omega}_\mu\cos\theta\sin\varphi\,\mathrm{I}.
        \end{split}
    \end{align}
\end{subequations}

Before the spontaneous symmetry breaking of the additional $U(1)_C$ symmetry, the field $ \mathit{\Omega}_\mu  $ similar as $ A_{\mu } $ must be coupled with a conserved lepton-quark current and the related charge should be conserved. Because quark axial currents are not conserved even on the massless level (triangle anomaly) then $ \mathit{\Omega}_\mu  $ must be coupled with a linear combination of the neutral quark vector currents which, as we known, are conserved separately for both ($u$ and $d$) quarks. If we analyze the processes with quark and lepton participation, and take into account the values of the charge $\Omega$ for leptons we conclude that only following possibility arises: the quark (antiquark) $u$ $(\overline{u})$ has the $\Omega$-charge value $-\frac{1}{3}\left(+\frac{1}{3}\right)$ while $d\left(\overline{d}\right)$ has this value equal to $+\frac{2}{3}\left(-\frac{2}{3}\right)$. Therefore the conserved quark current coupled to the field $\mathit{\Omega}_\mu$ takes the form $\frac13\overline{u}\gamma^{\mu }u-\frac23\overline{d}\gamma^{\mu }d$. Recall that the quark electric current has the form $-\frac23\overline{u}\gamma^{\mu }u+\frac13\overline{d}\gamma^{\mu }d$. 

Thus, by means of Eqs. \eqref{eq:Quark5} and \eqref{subeqn:Lepton4c}, the quark part of the extended Lagrangian takes the form
\begin{equation} \label{eq:Quark6}
    \begin{split}
        \mathcal{L}_{\mathrm{quark}} &= \overline{u}\gamma^{\mu }\partial_{\mu }i u+\overline{d} \gamma^{\mu }\partial_{\mu }i d
        -m_{u}\overline{u}u-m_{d}\overline{d}d+ \\
        &-eA_{\mu }J_Q^\mu-q\mathit{\Omega}_\mu J_\Omega^\mu
        -\frac{g}{2\sqrt{2}}\left(W_{\mu }^{+}J_{-}^{\mu }+W_{\mu }^{-}J_{+}^{\mu }\right)+ \\
        &-\frac{gZ_\mu}{2 \cos\theta \cos\varphi }\left(\left(g_{V}^{u} J_{u}^{\mu }-g_{A}^{u} J_{5u}^{\mu })+(g_{V}^{d} J_{d}^{\mu }-g_{A}^{d} J_{5d}^{\mu }\right)\right)
        -\frac{m_{u}}{\vartheta }\overline{u}Hu-\frac{m_{d}}{\vartheta }\overline{d}Hd,
    \end{split}
\end{equation}
where the quark currents are of the following form

\begin{equation} \label{eq:Quark7}
\begin{aligned} 
        J_{u}^{\mu }&=\overline{u} \gamma^{\mu }u, & \quad
        J_{5u}^{\mu }&=\overline{u} \gamma^{\mu }\gamma^{5}u, & \quad
        J_{+}^{\mu }&=\overline{d} \gamma^{\mu }\left(\mathrm{I}-\gamma^{5}\right)u, & \quad
        J_Q^\mu &=-\frac{2}{3}\overline{u}\gamma^{\mu }u+\frac13\overline{d}\gamma^{\mu }d, \\
        J_{d}^{\mu }&=\overline{d}\gamma^{\mu }d, & \quad
        J_{5d}^{\mu }&=\overline{d} \gamma^{\mu }\gamma^{5}d, & \quad
        J_{-}^{\mu }&=\overline{u} \gamma^{\mu }\left(\mathrm{I}-\gamma^{5}\right)d, & \quad
        J_\Omega^\mu &=\frac13\overline{u}\gamma^{\mu }u-\frac23\overline{d}\gamma^{\mu }d.
\end{aligned}
\end{equation}

The corresponding coupling constants $ g_{V/A}^{u/d} $ are given by the formulas:

\begin{equation} \label{eq:Quark8}
    \begin{aligned}
        g_{V}^{u} &= \frac{1}{2}\left(\left(1 -\frac{8}{3}\sin^{2}\theta\right)-\frac{4}{3}\sin^{2}\theta \sin^{2}\varphi\right), & \quad
        g_{A}^{u} &=\frac{1}{2},
    \\
        g_{V}^{d} &= \frac{1}{2}\left(\left(-1 +\frac{4}{3}\sin^{2}\theta\right)\cos^{2}\varphi +\frac{1}{3}\sin^{2}\varphi \right), & \quad
        g_{A}^{d} &= -\frac{1}{2}.
    \end{aligned}
\end{equation}

\noindent If $ \varphi \rightarrow 0 $ then $\mathcal{L}_{\mathrm{quark}}\rightarrow\mathcal{L}_{\mathrm{quark}}^\mathrm{SM} $. 

\section{A mirror symmetry of the model} \label{Sec:mirror}

The considered extension of the Standard Model has an apparent mirror symmetry of fundamental fermions. Indeed,  interchange of the fundamental leptons $\nu\leftrightarrow l$ and quarks $u\leftrightarrow d$, associated with the interchange of the gauge bosons   $\mathit{\Omega}_\mu\leftrightarrow A_\mu$, $Z_\mu\leftrightarrow Z_\mu$, $W_\mu^+\leftrightarrow W_\mu^-$  does not change the structure of the lepton and quark Lagrangians if associated with interchange of appropriate coupling constants $e\leftrightarrow q$,   $g_{V/A}^\nu\leftrightarrow g_{V/A}^l$, $g_{V/A}^u\leftrightarrow g_{V/A}^d$ and corresponding masses. Notice also, that the change  $\Omega\leftrightarrow \mathrm{Q}$ in the Eq.\eqref{eq:Intro1} does not change this formula. In each generation of fundamental fermions we have the following pattern of charges

\begin{table}[H]
\begin{center}
\begin{tabular}{c|c c|c c|c c|c c c c}
\hline
& \multicolumn{2}{l|}{leptons:} & \multicolumn{2}{l|}{quarks:} & \multicolumn{2}{l|}{baryons:} & \multicolumn{4}{l}{gauge bosons:} \\
& $\nu$ & $l$ & $u$ & $d$ & $n$ & $p$ & $A_\mu$ & $\mathit{\Omega}_\mu$ & $Z_\mu$ & $W_\mu^\pm$ \\
\hline
$\mathrm{Q}$ & $0$ & $-1$ & $+\sfrac23$ & $-\sfrac13$ & $0$ & $+1$ & $0$ & $0$ & $0$ & $\pm 1$ \\
$\Omega$ & $-1$ & $0$ & $-\sfrac13$ & +$\sfrac23$ & $+1$ & $0$ & $0$ & $0$ & $0$ & $\mp 1$ \\
\hline
\end{tabular}    
\end{center}
\caption{The patern of $\mathrm{Q}$ and $\Omega$ charges for extended SM particles.}
\end{table}
\noindent and with oposite charges for antiparticles.

Moreover, from the  Lagrangians \eqref{eq:Lepton5} and \eqref{eq:Quark6} we can deduce that the  $\Omega$ charge of gauge  fields $W_\mu^\pm$  is  $\mp1$ respectively, while $A_\mu$, $\mathit{\Omega}_\mu$ and $Z_\mu$ are neutral which is in agreement with Eq.\eqref{eq:Intro1}.

\section{Spontaneous symmetry breaking in the Higgs sector} \label{Sec:breakingHiggs}

In the standard minimal scheme of the spontaneous symmetry breaking the Goldstone--Higgs boson $ G $ is assumed to be the $ SU(2) $ iso-doublet. In the unitary gauge 
$ G\rightarrow\frac{1}{\sqrt{2}}\begin{pmatrix} 0 \\ H+\vartheta \\ \end{pmatrix} $. The generation of lepton and quark masses from the Yukawa type interaction of the elementary fermions with $ H $ field gives the Dirac masses for all elementary fermions (including neutrinos) as is put down in Eqs. \eqref{eq:Lepton5} and \eqref{eq:Quark6}). Because only three degrees of freedom of the Goldstone--Higgs iso-doublet can be absorbed by vector bosons as theirs third polarization, then only three gauge fields can acquire mass in this way, namely $ W_{\mu }^{\pm } $ and $ Z_{\mu } $. This means, that we have two possibilities: (a) generate a mass of the field $ \mathit{\Omega}_\mu  $ by means of the Higgs mechanism with an additional Goldstone--Higgs like field (or by the St{\"u}ckelberg mechanism \cite{Kors2004AStuckelberg}), (b) leave $ \mathit{\Omega}_\mu  $ as the massless field accepting its long-range character. However, irrespective of this question we should consider transformation properties of $ G $ under extended group transformations. Most general transformation of the iso-doublet $ G $, according to the extended electroweak group is of the form

\begin{equation} \label{eq:BreakHiggs1}
    G' = e^{\displaystyle i\frac{\boldsymbol{\alpha}\boldsymbol{\tau}}{2}}e^{\displaystyle i\frac{y_{H}\beta }{2}}e^{\displaystyle i\frac{c_{H}\gamma }{2}}G, 
\end{equation}

\noindent where $ y_{H} = 1 $ is the hypercharge of $ H $ while $ c_{H} $ fix a generator of an irreducible representation of the group $ U\left(1\right)_{C} $.

Notice, that $\overline{G} = i\tau_{2}G^{\ast } $ transforms as follows

\begin{equation} \label{eq:BreakHiggs2}
    \overline{G}^{'} = e^{\displaystyle i\frac{\boldsymbol{\alpha}\boldsymbol{\tau}}{2}}e^{\displaystyle -i\frac{y_{H}\beta }{2}}e^{\displaystyle -i\frac{c_{H}\gamma }{2}}\overline{G}. 
\end{equation}

The lepton and quark mass terms and related Yukawa interaction term with $ H $ in Eqs. \eqref{eq:Lepton5} and \eqref{eq:Quark6} arise in the unitary gauge from the invariant terms $ -\frac{\sqrt{2}m_{l}}{\vartheta }\overline{L}Gl_{R}-\frac{\sqrt{2}m_{\nu }}{\vartheta }\overline{L}\overline{G}\nu_{R}+h.c. $ for leptons and from analogous invariant terms for up and down quarks. Requirements of invariance of the terms above under transformations \eqref{eq:Prelim2}, \eqref{eq:Quark1}, \eqref{eq:BreakHiggs1} and \eqref{eq:BreakHiggs2} enforces conditions on the C-charges: $c_H$ and $c_i$, reduced to

\begin{equation} \label{eq:BreakHiggs3}
    c_{H} = c_{2}-c_{1}.
\end{equation}

Now, the covariant derivative of the Goldstone–Higgs field is given by\\ $iD_{\mu }G=\left(i\partial_{\mu }+\Gamma_{\mu }^{G}\right)G$ with

\begin{equation} \label{eq:BreakHiggs4}
    \begin{split}
        \Gamma_{\mu }^{G} &=\frac{g}{\sqrt{2}}\begin{pmatrix}
        0 & W_{\mu }^{+} \\ 
        W_{\mu }^{-} & 0 \\ 
        \end{pmatrix}+\frac{g}{2}\begin{pmatrix}
        W_{\mu }^{3} & 0 \\ 
        0 & -W_{\mu }^{3} \\ 
        \end{pmatrix}
        +\frac{y_{H}g'}{2}\begin{pmatrix}
        B_{\mu } & 0 \\ 
        0 & B_{\mu } \\ 
        \end{pmatrix}+\frac{c_{H}g''}{2}\begin{pmatrix}
        C_{\mu } & 0 \\ 
        0 & C_{\mu } \\ 
        \end{pmatrix}. 
    \end{split}
\end{equation}

Calculating the square of the covariant derivative of $G$ in the unitary gauge, we can identify the interaction terms of the Higgs field $H$ with the physical gauge bosons. We use the condition that the field $ H $ cannot interact with fields $ A_{\mu } $ and $ \mathit{\Omega}_\mu  $. The first condition realizes as identity while the second one gives relationship between $c_H$ and $c_1$

\begin{equation} \label{eq:BreakHiggs5}
    c_{1} = c_{H} \cos2\theta.
\end{equation}

Taking into account relations \eqref{eq:Lepton2}, \eqref{eq:Quark2}, \eqref{eq:Quark3}, \eqref{eq:Quark4}, \eqref{eq:BreakHiggs3} and \eqref{eq:BreakHiggs5}, rescaling $c_i\rightarrow c_i/c_H$ and $\tilde{c}_i\rightarrow \tilde{c}_i/c_H$ as well as $g''\rightarrow g''/c_H$ in the trasformation parameter $\gamma\rightarrow c_H \gamma$ (or equivalently puting $c_H=1$), we obtain explicit form of all C-charges, namely

\begin{equation} \label{eq:BreakHiggs6}
\begin{aligned}
    c_{1} &= \cos2\theta, & \quad
    c_{2} &=1+\cos2\theta, & \quad
    c_{3} &= -\left(1-\cos2\theta\right), \\
    \tilde{c}_{1} &= -\tfrac{1}{3}\cos2\theta, & \quad  
    \tilde{c}_{2} &=1-\tfrac{1}{3}\cos2\theta, & \quad
    \tilde{c}_{3} &= -\left(1+\tfrac{1}{3}\cos2\theta\right),
\end{aligned}
\end{equation}

\noindent with $c_H=1$, as well as the relationship between coupling constants $g$, $g'$ and $g''$

\begin{equation} \label{eq:BreakHiggs7}
\setlength\arraycolsep{12pt}
    \begin{aligned}
        g' = g\tan\theta, & \quad g'' = g\dfrac{\tan\varphi}{\cos\theta}.
    \end{aligned}
\end{equation}

Concluding, the C-charges $c_i$  and ${\tilde{c}}_i$ are determined by the parameter $\theta$ while the coupling constants $g'$ and   $g''$  by  $g$, angle $\theta$ and $\varphi$.

As one of consequences we obtain Higgs part of Lagrangian in the standard form 

\begin{equation} \label{eq:BreakHiggs8}
    \begin{split}
        \mathcal{L}_{\mathrm{HWZ}} &=\frac{1}{2}\partial_{\mu }H\partial^{\mu }H-\frac{1}{2}m_{H}^{2}H^{2}
        -\frac{1}{2\vartheta }m_{H}^{2}H^{3}\left(1+\frac{1}{4\vartheta }H\right)+
    \\
        &+\left(M_{W}^{2}W_{\mu }^{+}W^{\mu -}+\frac{1}{2}M_{Z}^{2}Z_{\mu }Z^{\mu }\right)\left(1+\frac{2}{\vartheta }H+\frac{1}{\vartheta^{2}}H^{2}\right),
    \end{split}
\end{equation}
but with 
\begin{equation} \label{eq:BreakHiggs9}
    \begin{aligned}
        M_{W} =\dfrac{g\vartheta }{2} \quad & \quad \textrm{and} & \quad M_{Z} =\dfrac{g\vartheta }{2 \cos \theta \cos \varphi}.
    \end{aligned}
\end{equation}

Therefore, a modified relationship between SM vector bosons masses arise
\begin{equation} \label{eq:BreakHiggs10}
    \frac{M_{W}}{M_{Z}} = \cos \theta \cos \varphi.
\end{equation}

Under a natural identification of the angle $ \theta $ with the Weinberg angle $ \theta_{W} $, the formula \eqref{eq:BreakHiggs10} tell us that the angle $ \varphi $ should be very small and a constraint on the upper bound of $ \varphi $ can be done from \eqref{eq:Lepton7}, \eqref{eq:Quark8} and \eqref{eq:BreakHiggs10}. Notice that the $ \Omega $ charge of the Higgs boson $ H $ is equal $ 0 $ because gauge boson $ \mathit{\Omega}_\mu  $ does not interact with the field $H$. 

\section{Spontaneous symmetry breaking in the dark sector} \label{Sec:breakingDark}

To avoid the mentioned in the introduction consequences related to the long range interaction of the field $ \mathit{\Omega}_\mu  $ we consider here a scenario with the spontaneous breaking of the additional $ U(1)_\Omega $ group with help of a complex iso-singlet Higgs field $ \chi $ interacting with the vector field $\mathit{\Omega}_\mu$ only. By applying the standard procedure we obtain the corresponding Lagrangian in the form 
\begin{equation} \label{eq:BreakDark1}
    \begin{split}
        \mathcal{L}_{\chi \Omega } &= \frac{1}{2}\partial_{\mu }\chi \partial^{\mu }\chi -\frac{1}{2} m_{\chi }^{2}\chi^{2}
        -\frac{1}{2}q\frac{ m_{\chi }^{2}}{M_\mathit{\Omega}} \chi^{3}- \frac{1}{8}q^{2}\left(\frac{ m_{\chi } }{M_\mathit{\Omega}}\right)^{2} \chi^{4}+
    \\
        &+\frac12 M_\mathit{\Omega}^2\mathit{\Omega}_\mu\Omega^\mu\left(1+\frac{2q}{M_\mathit{\Omega}}\chi+\left(\frac{q}{M_\mathit{\Omega}}\right)^2\chi^2\right),
    \end{split}
\end{equation}
where $m_\chi$ is the mass of the Higgs $\chi$ particle and $M_\mathit{\Omega}$ is the mass of the gauge boson $\mathit{\Omega}_\mu$. In this case, the VEV of the Higgs-Goldstone boson is given by the formula $\vartheta_\chi=M_\mathit{\Omega}/q$. Notably, the field $\chi$, aside from self-interaction, interact exclusively with $\mathit{\Omega}_\mu$. For this reason $\chi$ is considered a viable  dark matter (DM) particle candidate (dark Higgs). However, this is only feasible if the particles are cosmologically stable. In the standard scenario where $m_\chi>M_\mathit{\Omega}$ this stability can be achieved by ensuring an extremely small $q$ which suppress decays of $ \chi $ into pair of vector bosons $ \mathit{\Omega}_\mu  $ or with a special structure of the interaction term which forbids decays of $\chi$. However, a more natural and simpler assumption is that the mass $m_\chi$ of the Higgs particle $\chi$ is smaller than the mass $ M_\mathit{\Omega} $ of the vector boson $ \mathit{\Omega}_\mu  $. Thus we assume that $ M_\mathit{\Omega}> m_{\chi } $. In Sec.\ref{Sec:stability} we return to the DM  stability question in our model. The gauge boson $\mathit{\Omega}_\mu$ interacts with $\chi$ but, as follows from Lagrangian \eqref{eq:Lepton5} and  \eqref{eq:Quark6}, it also interacts with elementary fermions making it a mediator particle between the dark and SM sector. Considering that $q=\cos{\theta}\sin{\varphi}$ the coupling $q$ must be smaller than the weak coupling constant $g$. To maintain consistency between the Standard Model's predictions, this extended model (see especially Eqs. \eqref{eq:Lepton5}, \eqref{eq:Quark6}, \eqref{eq:BreakHiggs10}) a small upper limit for $q$ is anticipated, which is typical for feebly interacting massive dark matter particles (FIMP). This issue will be explored in the next section. 

\section{Gauge fields Lagrangian; absence of anomalies} \label{Sec:gaugeL}

The gauge part of the full Lagrangian (except of the mass terms discussed below) is obtained from the manifestly gauge covariant Lagrangian

\begin{equation} \label{eq:GaugeL1}
\mathcal{L}_{\mathrm{gauge}} = -\frac{1}{4}\bm{W}_{\mu \nu }\bm{W}^{\mu \nu }-\frac{1}{4}B_{\mu \nu }B^{\mu \nu }-\frac{1}{4}C_{\mu \nu }C^{\mu \nu },
\end{equation}
by replacement in the tensors 

\begin{subequations} \label{eq:GaugeL2}
    \begin{align}
        \label{subeqn:GaugeL2a}
        W_{\mu \nu }^{i} &= \partial_{\mu }W_{\nu }^{i}-\partial_{\nu }W_{\mu }^{i}+g \varepsilon^{ijk} W_{\mu }^{j}W_{\nu }^{k},    
    \\
        \label{subeqn:GaugeL2b}
    B_{\mu \nu } &= \partial_{\mu }B_{\nu }-\partial_{\nu }B_{\mu },
    \\
        \label{subeqn:GaugeL2c}
    C_{\mu \nu } &= \partial_{\mu }C_{\nu }-\partial_{\nu }C_{\mu },
    \end{align}
\end{subequations}

\noindent the fields $W_\mu,B_\mu,C_\mu$ into $ W_{\mu }^{\pm } $, $ A_{\mu },Z_{\mu }, $ $ \mathit{\Omega}_\mu  $ according to the eqs. \eqref{eq:Lepton2}. The explicit form of the Lagrangian \eqref{eq:GaugeL1} is given in the Eq.\eqref{eq:A1}. A mathematical consistency of the model demands the cancellation of possible gauge and gravitational anomalies \cite{Peskin1995AnIntroduction} for three generations of elementary fermions. Its electroweak group differs from the SM group by factor $ U(1)_{C} $. To prove consistency of the model we should check only the anomaly cancellation conditions for triangle graphs of the type $ SU(3)_{C}^{2}\times U(1)_{C} $, $ SU(2)_{L}^{2}\times U(1)_{C} $, $ U(1)_{Y}^{2}\times U(1)_{C} $, $ U(1)_{Y}\times U(1)_{C}^{2} $, $ U(1)_{C}^{3} $, and for the gravitational anomaly for $ U(1)_{C} $. This can be easily done in the three generations context by using the connection forms \eqref{eq:Lepton4} and \eqref{eq:Quark5} and values of lepton and quark charges. 

The cancellation of anomalies in our model can be also immediately proved by the use of the set of conditions given in Ref.\cite{Applequist2003Nonexotic} (see also \cite{Ekstedt2018Minimal}, Eqs. 3.1, 3.2). Under the requirement of generation-independent charge assignment and using Eq.\eqref{eq:B1} they reduce to the following system:
\begin{equation} \label{eq:GaugeL3}
\begin{gathered}
    c_H=c_1-c_3=c_2-c_1=\tilde{c}_2-\tilde{c}_1=\tilde{c}_1-\tilde{c}_3=1, \\
    c_1=-3\tilde{c}_1, \quad  
    c_3=-2\tilde{c}_1-\tilde{c}_2, \quad
    c_2=-4\tilde{c}_1+\tilde{c}_2.
\end{gathered}    
\end{equation}
We can immediately check that the above system is equivalent to the system of Eqs. \eqref{eq:Lepton3}, \eqref{eq:Quark3}, \eqref{eq:BreakHiggs3}. As a result of anomaly cancellation, we can conclude that our model is free of anomalies.

\section{Dark matter freeze-in} \label{sec:freeze}

Dark matter is currently the most natural candidate for explaining the baryonic matter deficit observed in the description of our universe on cosmic scales \cite{Clowe2006ADirect,Roszkowski2018Wimp}. Therefore, the question of its evolution and abundance is especially important. If we correctly identify  $\chi$ particles as dark matter, their cosmic evolution should lead to the observed large relic abundance \cite{Navas2024Review}. There are two main mechanisms explaining the emergence of dark matter particles from the primordial plasma and their production with relic abundance. Both mechanisms are described by an appropriate system of Boltzmann equations in the Friedmann–Lemaître–Robertson–Walker (FLRW) background \cite{Kolb2018TheEarly,Gondolo1991Cosmic,Lee1977Cosmological}. 

The first mechanism, the freeze-out scenario \cite{Burgess2001TheMinimal,Roszkowski2018Wimp,Steigman2012Precise,Baer2015Dark,Heikinheimo2017Wimp}, assumes that at sufficiently high temperatures, interactions between dark matter (DM) particles and standard model (SM) particles are strong enough to achieve thermal equilibrium with the primordial plasma. As the temperature drops below the particle mass, the annihilation process slows down compared to the expansion rate of the universe. As a result, this leads to the freeze-out of the DM species, and the comoving number density of dark matter particles becomes constant. 

The second mechanism, the freeze-in process \cite{Hall2010Freeze,Bernal2017TheDawn,Belanger2020Dark,Du2022Revisiting}, is governed by very weak couplings between dark matter particles and the SM plasma, meaning that the two populations never reach thermal equilibrium. According to the freeze-in scenario, the DM population starts with out-of-equilibrium particles of extremely low number density, which gradually accumulate, causing their density to increase. After the intensity of creation and annihilation interactions diminishes, dark matter particles freeze in without ever reaching equilibrium with the primordial plasma. As a result, the evolution of the frozen-in dark matter population should yield the observed relic abundance.  
Arguments presented in Sec.\ref{Sec:breakingDark} lead to the conclusion that the freeze-in mechanism is the more likely process for $\chi$ dark matter generation. In our case, dark matter particles $\chi$ interact with the SM sector \emph{via} a mediator particle $\mathit{\Omega}_\mu$ which couples to SM fermions through interactions with quarks and neutrinos (Eqs. \eqref{eq:Lepton5},\eqref{eq:Quark6} and \eqref{eq:BreakDark1}). From here on, we use the abbreviation $\mathit{\Omega}$ for the gauge vector bosons $\mathit{\Omega}_\mu$ and $f$ for the elementary fermions of the Standard Model. 

The reaction related to the creation and annihilation of dark matter has the form: 
\begin{subequations} \label{eq:DMFreeze1}
    \begin{equation} \label{subeqn:DMFreeze1a}
        \mathit{\Omega}+\mathit{\Omega}\longleftrightarrow\chi +\chi,
    \end{equation}
and is associated with creation and annihilation processes:
    \begin{equation} \label{subeqn:DMFreeze1b}
        \mathit{\Omega}+\mathit{\Omega}\longleftrightarrow f+\bar{f},
    \end{equation}
and the decay / inverse decay processes:
    \begin{equation} \label{subeqn:DMFreeze1c}
        \mathit{\Omega}\longleftrightarrow f+\bar{f}.
    \end{equation}
\end{subequations}

\noindent These processes lead to the creation (or annihilation) of $\mathit{\Omega}_\mu$ particles by SM fermions and the creation (or annihilation) of $\chi$  particles.

By means of the Feynman diagrams deduced from the Lagrangian \eqref{eq:BreakDark1}, at tree level, in the unitary gauge, amplitude of the process \eqref{subeqn:DMFreeze1a} has the form given in Fig.\ref{fig:FeynmanOOxx}.

\begin{center}
\begin{figure}[H]
\[
\begin{aligned}
\mathbb{M}&=
\begin{tikzpicture} [baseline=(a.base)]
\begin{feynman}
\vertex[dot] (a) {};
\vertex [above left=of a] (i1) {\(k\)};
\vertex [below left=of a] (i2) {\(p\)};
\vertex [above right=of a] (f1) {\(k_\chi\)};
\vertex [below right=of a] (f2) {\(p_\chi\)};
\diagram* [horizontal=i1 to f1] { 
  (i1) -- [draw=none] (i2),
  (i1) -- [boson] (a),
  (i2) -- [boson] (a),
  (a)-- [scalar] (f1), 
  (a)-- [scalar] (f2), 
}; 
\vertex [below=0.75 of a] {\(2iq^2\eta_{\mu\nu}\)};
\end{feynman}
\end{tikzpicture}
+
\begin{tikzpicture} [baseline=(a.base)]
\begin{feynman}
\vertex[dot] (a) {};
\vertex [above left=of a] (i1) {\(k\)};
\vertex [below left=of a] (i2) {\(p\)};
\vertex[dot] [right=of a] (b) {};
\vertex [above right=of b] (f1) {\(k_\chi\)};
\vertex [below right=of b] (f2) {\(p_\chi\)};
\diagram* [vertical'=a to b,horizontal=i1 to a] { 
  (i1) -- [draw=none] (i2),
  (i1) -- [boson] (a),
  (i2) -- [boson] (a),
  (a) -- [scalar] (b),
  (b)-- [scalar] (f1), 
  (b)-- [scalar] (f2), 
}; 
\vertex [left=1 of a] {\(2iqM_\mathit{\Omega}\eta_{\mu\nu}\)};
\vertex [right=0.75 of b] {\(3iq\frac{m_\chi^2}{M_\mathit{\Omega}}\)};
\vertex [right=0.75 of a] (c);
\vertex [above=0.25 of c] {\(\frac{i}{\left(k+p\right)^2-m_\chi^2}\)};
\end{feynman}
\end{tikzpicture}
+\\
&+
\begin{tikzpicture} [baseline=(center.base)]
\begin{feynman}
\vertex at (0,0) (center);
\vertex at (0,1) (i1) {\(k\)};
\vertex at (0,-1) (i2) {\(p\)};
\vertex[dot] [right=of i1] (a) {};
\vertex[dot] [right=of i2] (b) {};
\vertex [right=of a] (f1) {\(k_\chi\)};
\vertex [right=of b] (f2) {\(p_\chi\)};
\diagram* [vertical'=a to b,horizontal=i1 to a] { 
  (i1) -- [boson] (a),
  (i2) -- [boson] (b),
  (a) -- [boson, edge label=\(\frac{-i}{l^2-M_\mathit{\Omega}^2}\left(\eta_{\mu\nu}-\frac{l_\mu l_\nu}{M_\mathit{\Omega}^2}\right)\)] (b),
  (a)-- [scalar] (f1), 
  (b)-- [scalar] (f2), 
}; 
\vertex [above=0.375 of a] {\(2iqM_\mathit{\Omega}\eta_{\mu\nu}\)};
\vertex [below=0.375 of b] {\(2iqM_\mathit{\Omega}\eta_{\mu\nu}\)};
\end{feynman}
\end{tikzpicture}
+
\begin{tikzpicture} [baseline=(center.base)]
\begin{feynman}
\vertex at (0,0) (center);
\vertex at (0,1) (i1) {\(k\)};
\vertex at (0,-1) (i2) {\(p\)};
\vertex[dot] [right=of i1] (a) {};
\vertex[dot] [right=of i2] (b) {};
\vertex [right=of a] (f1) {\(k_\chi\)};
\vertex [right=of b] (f2) {\(p_\chi\)};
\diagram* [vertical'=a to b,horizontal=i1 to a] { 
  (i1) -- [boson] (a),
  (i2) -- [boson] (b),
  (a) -- [boson] (b),
  (a)-- [scalar] (f2), 
  (b)-- [scalar] (f1), 
}; 
\vertex [above=0.375 of a] {\(2iqM_\mathit{\Omega}\eta_{\mu\nu}\)};
\vertex [below=0.375 of b] {\(2iqM_\mathit{\Omega}\eta_{\mu\nu}\)};
\vertex [below=1 of a] (c);
\vertex [left=0 of c] {\(\frac{-i}{r^2-M_\mathit{\Omega}^2}\left(\eta_{\mu\nu}-\frac{r_\mu r_\nu}{M_\mathit{\Omega}^2}\right)\)};
\end{feynman}
\end{tikzpicture}
\end{aligned}
\]
where $l=k_\chi-k$, $r=p_\chi-k$
\caption[]{Amplitude of the $\mathit{\Omega+\Omega}\leftrightarrow\chi+\chi$ process at tree level.}
\label{fig:FeynmanOOxx}
\end{figure}
\end{center}

Following the standard procedure  \cite{Kolb2018TheEarly,Gondolo1991Cosmic}, we use process  \eqref{subeqn:DMFreeze1a} to investigate the Hubble-covariant evolution of dark matter number density. We also calculate the density evolution of $\mathit{\Omega}$ bosons using reactions \eqref{eq:DMFreeze1}. 
The evolution of dark matter particles $\chi$ and mediator gauge bosons $\mathit{\Omega}$ is driven by the competition between the production and annihilation of these particles, as described by the Lee-Weinberg modification of the Boltzmann equation \cite{Kolb2018TheEarly,Lee1977Cosmological}

\begin{equation} \label{eq:DMFreeze2}
    \frac{dn}{dt}+3Hn = \left\langle \sigma v_\mathrm{\text{M{\o}l}} \right\rangle \left(n_{eq}^{2}-n^{2}\right).
\end{equation}
Here $n$ is the number density of $\chi$ particles while $n_{eq}$ is their Boltzmann equilibrium number density. The Hubble parameter $ H = a^{-1}\frac{da}{dt} $, where $a$ is the cosmological (FLRW) scale factor. The symbol $\left\langle \sigma v_\mathrm{\text{M{\o}l}} \right\rangle $ denotes the thermally \text{M{\o}l}ler averaged cross section \cite{Gondolo1991Cosmic}. In our case $\left\langle \sigma v_\mathrm{\text{M{\o}l}} \right\rangle $ takes the form   
  
\begin{equation} \label{eq:DMFreeze3}
    \left\langle \sigma v_\mathrm{\text{M{\o}l}} \right\rangle_{\chi\chi\leftrightarrow \Omega\Omega} =\frac{1}{8m_{\chi } TK_{2}^{2}\left(\dfrac{m_{\chi }}{T}\right) }\int_{4M_\mathit{\Omega}^{2}}^{\infty }ds\sqrt{s}\left(s-4m_{\chi }^{2}\right)K_{1}\left(\frac{\sqrt{s}}{T}\right)\sigma \left(q,s,m_{\chi },M_\mathit{\Omega}\right),
\end{equation}

\noindent where the cross-section $\sigma\left(q,s,m_\chi,M_\mathit{\Omega}\right)$ is given in Appendix \ref{Sec:csOOxx}, $T$ is the primordial plasma temperature, $ K_{1} $ and $ K_{2} $ are the modified Bessel functions of the first and second kind respectively. In the following, it will be convenient to use the variable $ x = m_{\chi }/T $ instead of $ T $. In terms of the parameter $x$ the FLRW Hubble covariant dark matter densities are defined as $Y(x)=n(x) / s(x)$ and  $Y_{eq}(x)=n_{eq}(x) / s(x)$, where the Boltzmann equilibrium number density is given by
\begin{equation} \label{eq:DMFreeze4}
    n_{eq}\left(x\right) =\frac{1}{2\pi^{2}}\frac{m_{\chi }^{2}}{x}K_{2}(x),
\end{equation}
while the temperature-dependent entropy of the universe takes the form
\begin{equation} \label{eq:DMFreeze5}
    s(x) =\frac{2\pi^{2}}{45}\frac{g_*\left(\frac{m_{\chi }}{x}\right) m_{\chi }^{3}}{x^{3}}.
\end{equation}

\begin{figure}[H]
    \centering
    \includegraphics[width=0.667\textwidth]{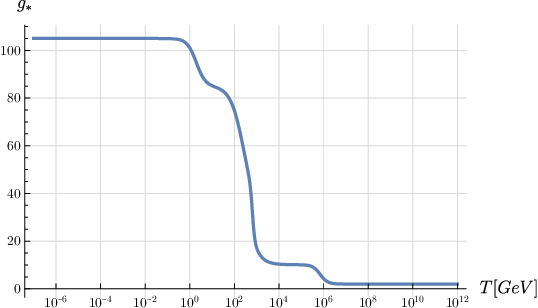}
    \caption[Effective degrees of fredom of the primordial plasma]{The function $g_*(T)\equiv g_*\left(x,m_\chi\right)$, for $x=m_\chi/T$, describing effective, thermally coupled discrete relativistic degrees of freedom of the primordial plasma \cite{Steigman2012Precise,Laine2006Quark}}
    \label{fig:df}
\end{figure}

The function $g_*(T)$, describing effective, thermally coupled discrete relativistic degrees of freedom of the primordial plasma is presented in Fig.\ref{fig:df}.

During the radiation era (the time of a dynamical evolution of DM particle density) the Hubble parameter $ H(x) $ is given by the formula 
\begin{equation} \label{eq:DMFreeze6}
    H\left(x\right) =\frac{\pi\sqrt{g_*\left(\dfrac{m_{\chi }}{x}\right) } m_\chi^{2}}{3\sqrt{10 }x^{2}M_{P}},
\end{equation}
where $ M_{P} = 2.435\cdot 10^{18} \hspace{1ex} GeV $ is the reduced Planck mass. Then the Boltzmann Eq.\eqref{eq:DMFreeze2} takes the final covariant form of the Riccati type equation 

\begin{equation} \label{eq:DMFreeze7}
    \frac{dY_\chi (x)}{dx} =\frac{s(x)\left\langle \sigma v_\mathrm{\text{M{\o}l}} \right\rangle_{\chi\chi\leftrightarrow \Omega\Omega} }{x H(x)}\left(Y_{\chi eq}^{2}(x)-Y_\chi^{2}(x)\right),
\end{equation}
where
\begin{equation} \label{eq:DMFreeze8}
   Y_{\chi eq}(x) =\frac{45}{4\pi^4}\frac{x^2}{g_*\left(\dfrac{m_\chi}{x} \right)}K_2(x).
\end{equation}

The equation \eqref{eq:DMFreeze7} can be solved by numerical methods under typical  for freeze-in initial condition $Y_\chi\left(x_0\right)=10^{-18}$ with $x_0\ll 1$, which means that initial DM abundance was very small. Our aim is to obtain, as evolution effect, the observed dark matter abundance. Notice firstly that the present epoch value of $x$ is given by $x_p=m_\chi / T_p$ where the CMBR temperature $T_p=2.725\ K=2.35\cdot 10^{-13}\ GeV$  so
\begin{equation} \label{eq:DMFreeze9}
    x_p =\frac{m_\chi}{2.35\cdot 10^{-13}\ GeV}.
\end{equation}
    
To obtain a physical solution of the Eq.\eqref{eq:DMFreeze7}, we should compare the obtained covariant density $Y\left(x_p\right)$  with the measured dark matter parameter density $\Omega_{DM} h^2 =$\\$=m_{DM}n_{DM}\left(T_p\right)h^2/\rho_c=0.11862$  \cite{Navas2024Review},  where the critical density $\rho_c=8\cdot10^{-47} h^2\ GeV^4$ and $h=H\left(x_p\right)\ Mpc\cdot s/100\ km$. Using the form \eqref{eq:DMFreeze5} of the entropy of universe we obtain, after identification $\chi=DM$, that

\begin{equation} \label{eq:DMFreeze10}
    Y_{DM}\left(x_p\right)=Y_{\chi}\left(x_p\right)\approx\frac{8.36}{m_\chi}\cdot 10^{-10}\ GeV.
\end{equation}

\begin{figure}[H]
    \centering
    \includegraphics[width=0.667\textwidth]{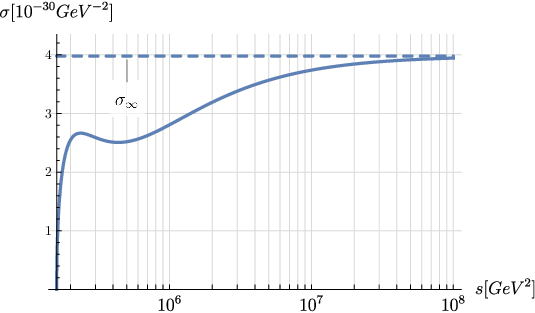}
    \caption[The cross section of the process $\mathit{\Omega}+\mathit{\Omega}\leftrightarrow \chi+\chi$]{The cross section $\sigma\left(q,s,m_\chi,M_\mathit{\Omega}\right)$ of the process $\mathit{\Omega}+\mathit{\Omega}\leftrightarrow \chi+\chi$ as function of the Mandelstam variable $s$ for $q=10^{-6}$, $m_\chi=150$, $M_\mathit{\Omega}=200$. The limiting value of $\sigma$ for $s\rightarrow\infty$, $\sigma_\infty=q^4 / 2\pi M_\mathit{\Omega}^2$.}
    \label{fig:csOOxx}
\end{figure}

Numerical solutions of equation \eqref{eq:DMFreeze7} for the $\chi$ population can be obtained by using the  cross section $\sigma(q,s,m_\chi,M_\mathit{\Omega} )$,  which analytic form is given in the Appendix \ref{Sec:csOOxx}. Its plot for a specific values of  the coupling constant $q$ and masses $m_\chi$  and  $M_\mathit{\Omega}$ is given in Fig.\ref{fig:csOOxx}.

\begin{figure}[H]
    \centering
    \includegraphics[width=0.667\textwidth]{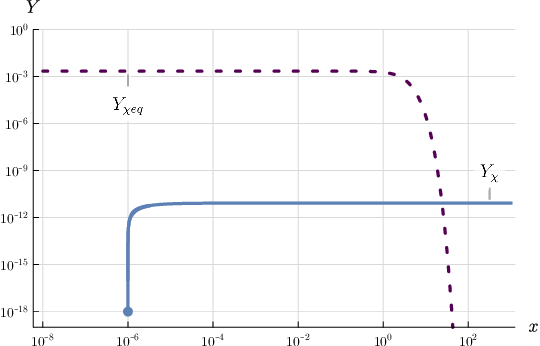}
    \caption[]{Numerical solution of the equation \eqref{eq:DMFreeze7} for the specific values: $q=10^{-7},\ m_\chi=100\ GeV,\ M_\mathit{\Omega}=174.5\ GeV,\  x_0=10^{-6},\  Y_\chi (x_0 )=10^{-18}$. The dashed purple line denotes the equilibrium trajectory $Y_{eq}$. The  covariant density $Y_\chi (x)$ (blue line) starts in $x_0=10^{-6}$ having very small initial density $10^{-18}$ with a very rapid growth and next evolve into plateau. After passing through the cross point $Y_\chi (x_c)=Y_{\chi eq} (x_c)$, the density $Y_\chi (x)=\mathrm{const}$.}
    \label{fig:dens_chi}
\end{figure}

With use of Eqs. \eqref{eq:DMFreeze3} and \eqref{eq:A2} we can calculate and analyze variety of numerical solutions of Eq.\eqref{eq:DMFreeze7} under different choices of parameters $q,m_\chi,M_\mathit{\Omega}$ and initial conditions $Y_{\chi 0}=Y_\chi (x_0)$ assuming the DM condition \eqref{eq:DMFreeze10}. A typical solution of \eqref{eq:DMFreeze7}  for $q\ll 1$, $1\ GeV<m_\chi<M_\mathit{\Omega}<1\ PeV$ and $x_0<1$ has the form presented in the Fig.\ref{fig:dens_chi}. 

It is important to stress that values of parameters $q, x_0, m_\chi, M_\mathit{\Omega}$ are not arbitrary but they are restricted by the condition \eqref{eq:DMFreeze10} $Y_{DM} (x_p )=Y_\chi (x_p )$. Indeed, from the Fig.\ref{fig:dens_chi} we see that for $x>x_c$, where $x_c$  is the $x$ coordinate  of  cross point of the $Y_\chi$   and $Y_{\chi eq}$  trajectories, a sudden decreasing the Boltzmann equilibrium density $Y_{\chi eq}$ take place,  while the density $Y_\chi (x)$ leaves unchanged. This  implies  that for $x>x_c$ Eq.\eqref{eq:DMFreeze7}  reduces to the form
\begin{equation} \label{eq:DMFreeze11}
    \frac{dY_\chi(x)}{dx}\approx -\frac{s(x)\left\langle \sigma v_{\text{M{\o}l}}\right\rangle}{xH(x)}Y_\chi^2(x).
\end{equation}
Eq.\eqref{eq:DMFreeze11} is easy to integrate in the interval $\left(x_c,x\right)$; its solution is of the form 
\begin{equation} \label{eq:DMFreeze12}
    Y_\chi(x)=\frac{Y_\chi(x_c)}{1+Y_\chi(x_c)\displaystyle\int_{x_c}^x{dx\dfrac{s(x)\left\langle\sigma v_{\text{M{\o}l}}\right\rangle}{xH(x)}}}.
\end{equation}
Taking into account that the second term in the denominator in Eq.\eqref{eq:DMFreeze12} is extremely small, we conclude that for $x>x_c$ with a very good approximation
\begin{equation} \label{eq:DMFreeze13}
    Y_\chi(x)\approx Y_\chi(x_c)=\mathrm{const}.
\end{equation}

\noindent This leads to a very important equality
\begin{equation} \label{eq:DMFreeze14}
    Y_\chi(x_c)\approx Y_\chi(x_p)\approx\frac{8.36}{m_\chi}\cdot 10^{-10}\ GeV,
\end{equation}
confirmed by numerical calculations. The value of  $x_c$ can be calculated from the equality $Y_\chi (x_c )= Y_{\chi eq} (x_c)$ and e.g. for  $m_\chi<1\ TeV$ the inequality $21<x_c<26$  holds.   The result \eqref{eq:DMFreeze13} tells us that the covariant number density of frozen-in population $\chi$ stabilizes (plateau in Fig.\ref{fig:dens_chi}) as well as the condition \eqref{eq:DMFreeze14} should hold. In the next section, we will discuss the simultaneous thermal evolution of both populations $\chi$ and $\mathit{\Omega}$.

\section{Thermal evolution of dark matter and mediator population} \label{Sec:evolution}

Numerical analysis of the Eq.\eqref{eq:DMFreeze7} under the condition \eqref{eq:DMFreeze14} leads to a useful relationship between the parameters of the model. Namely, we obtain a very accurate formula respecting equality \eqref{eq:DMFreeze14}
\begin{equation} \label{eq:Thermal1}
    \sqrt{\frac{x_0}{\omega}}\approx 1.745\cdot 10^{11}q^2,
\end{equation}
where $\omega\equiv \left(m_\chi/M_\mathit{\Omega}\right)^2$.

The formula \eqref{eq:Thermal1} is at least very well satisfied in the range of parameters  $10\ GeV<m_\chi<10\ TeV$ and  $10^{-10}<x_0<10^{-1}$. In particular, values of $q, m_\chi, M_\mathit{\Omega}$ and  $x_0$ in Fig.\ref{fig:dens_chi} satisfy the relation \eqref{eq:Thermal1}. Because $x_0=m_\chi/T_0$, where $T_0$ is the temperature at the beginning of the $\chi$ particles evolution, then from  \eqref{eq:Thermal1} it follows that
\begin{equation} \label{eq:Thermal2}
    T_0=0.3284\cdot 10^{-22}  \frac{m_\chi}{\omega q^4}.
\end{equation}

\noindent Notice, that $m_\chi=100\ GeV$ and $x_0=10^{-10}$ correspond to temperature $T_0=10^{12}\ GeV$ considered as the maximal temperature of the radiation-dominated era. On the other hand, the point $x_0=1$, critical for a thermodynamical equilibrium between matter and radiation, defines the temperature $T_\mathrm{crit}=m_\chi$, which can be treated as a minimal temperature for the beginning of DM creation. Taking  the above  into account, we can determine a very conservative range of the coupling constant $q$, namely

\begin{equation} \label{eq:Thermal3}
    2.394\cdot 10^{-9} \left(\frac{m_\chi}{\omega\,1\,GeV}\right)^{\frac14}<q<2.394\cdot 10^{-6}\frac{1}{\omega^\frac14}.
\end{equation}

\noindent In the Fig.\ref{fig:dens_chi} we used  $\omega=0.328$  and $x_0=10^{-6}$  as well as $q=10^{-7}$.

Now, to achieve a clear picture of the thermal evolution of dark matter, we should supplement the density evolution determined by the equation \eqref{eq:DMFreeze7} with the evolution of the $\mathit{\Omega}$ boson population density related to the processes \eqref{eq:DMFreeze1}. The corresponding Boltzmann equation describing evolution of the covariant  density $Y_\mathit{\Omega} (x)$ takes the form

\begin{equation} \label{eq:Thermal4}
\begin{split}
    \frac{dY_\mathit{\Omega}(x)}{dx}&=\frac{s(x)}{xH(x)}\left(\left\langle\sigma v_{\text{M{\o}l}}\right\rangle_{\mathit{\Omega\Omega}\leftrightarrow\chi\chi}+\left\langle\sigma v_{\text{M{\o}l}}\right\rangle_{\mathit{\Omega\Omega}\leftrightarrow f\bar{f}}\right)\left(Y_{\Omega eq}^2(x)-Y_\Omega^2(x)\right)+\\
    &+\frac{\langle\Gamma\rangle_{\mathit{\Omega}\leftrightarrow f\bar{f}}}{xH(x)}\left(Y_{\Omega eq}(x)-Y_\Omega(x)\right),
\end{split}
\end{equation}

\noindent where the equilibrium density of the $\mathit{\Omega}$ bosons with the SM plasma is given by the formula

\begin{equation} \label{eq:Thermal5}
    Y_{\mathit{\Omega} \mathrm{eq}}(x)=\frac{45}{4\pi^4}\frac{M_\mathit{\Omega}^2x^2}{m_\chi^2g\left(\frac{x}{m_\chi}\right)}K_2\left(x\frac{M_\mathit{\Omega}}{m_\chi}\right).
\end{equation}

\begin{figure}[H]
    \centering
    \includegraphics[width=0.667\textwidth]{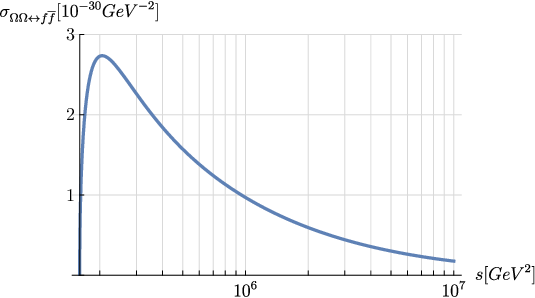}
    \caption[]{The cross section of the process $\mathit{\Omega+\Omega}\leftrightarrow f+\bar{f}$ as function of $s$ for coupling constant value $q=10^{-6}$ and masses $m_\chi=100\ GeV$ and $M_\mathit{\Omega}=174.5\ GeV$.}
    \label{fig:csOOff}
\end{figure}

Here $\left\langle\sigma v_{\text{M{\o}l}}\right\rangle_{\mathit{\Omega\Omega}\leftrightarrow\chi\chi}$ is  the \text{M{\o}l}ler thermally averaged cross section of the process \eqref{subeqn:DMFreeze1a} but calculated for the $\mathit{\Omega}$ boson density according to the rule \eqref{eq:DMFreeze3} Furthermore, $\left\langle\sigma v_{\text{M{\o}l}}\right\rangle_{\mathit{\Omega\Omega}\leftrightarrow f\bar{f}}$  is the total thermal cross section of all admissible SM fundamental fermion processes  \eqref{subeqn:DMFreeze1b}. The cross section of the process $\mathit{\Omega+\Omega}\leftrightarrow f+\bar{f}$ is given in the Appendix \ref{Sec:csOOff} while its plot is presented in the Fig.\ref{fig:csOOff}. 

The decay rate $\Gamma_\mathit{\Omega}$ of the process $\mathit{\Omega}\leftrightarrow f+\bar{f}$, given in Eq.\eqref{eq:A4}, leads to the thermally averaged decay/inverse decay  width $\langle\Gamma\rangle_{\mathit{\Omega}\leftrightarrow f\bar{f}}$ of the form

\begin{equation} \label{eq:Thermal6}
    \langle\Gamma\rangle_{\mathit{\Omega}\leftrightarrow f\bar{f}}=\frac{\gamma_\mathit{\Omega}\left(M_\mathit{\Omega}\right)q^2M_\mathit{\Omega}}{12\pi}\frac{K_1\left(x\dfrac{M_\mathit{\Omega}}{m_\chi}\right)}{K_2\left(x\dfrac{M_\mathit{\Omega}}{m_\chi}\right)},
\end{equation}
\noindent (see e.g. \cite{Frumkin2023Thermal}) where $\gamma_\mathit{\Omega}\left(M_\mathit{\Omega}\right)$ is the coefficient depending on quark masses entering the process  $\mathit{\Omega}\leftrightarrow f+\bar{f}$. Its plot as well as derivation is provided in \ref{Sec:dwO}.

\begin{figure}[H]
\centering
\setkeys{Gin}{width=\linewidth}
\begin{subfigure}{0.5\textwidth}
\includegraphics{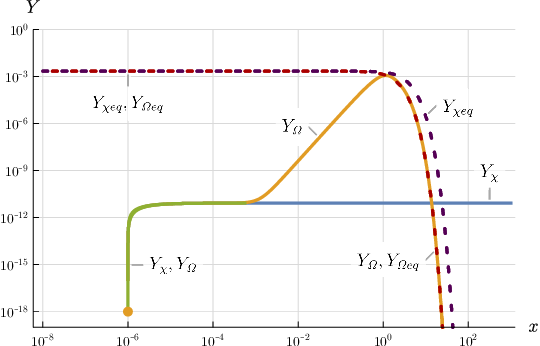}
\caption{$x_0=10^{-6}$}
\label{fig:evol_a}
\end{subfigure}
\hfil
\begin{subfigure}{0.5\textwidth}
\includegraphics{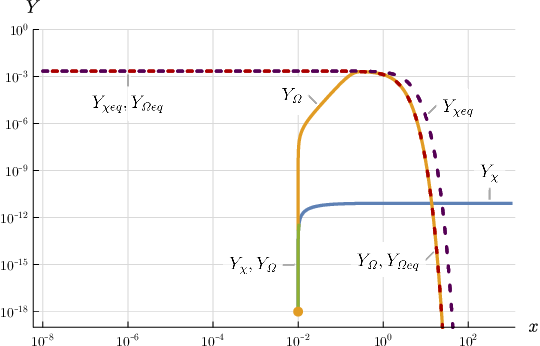}
\caption{$x_0=10^{-2}$}
\label{fig:evol_b}
\end{subfigure}
\caption{Evolution of the covariant density $Y_\chi (x)$ (green and blue line) and $Y_\mathit{\Omega} (x)$ (green and orange line) under assumption of the initial condition $Y_\chi \left( x_0\right) =Y_\mathit{\Omega} \left(x_0\right)=10^{-18}$ for values $x_0=10^{-6}$ (Fig.\ref{fig:evol_a}) and $x_0=10^{-2}$ (Fig.\ref{fig:evol_b}). The purple and red dashed lines represent evolution of the equilibrium densities $Y_{\chi eq}$ and $Y_{\mathit{\Omega} eq}$ respectively. The masses in the two configurations are the same i.e. $m_\chi=100\ GeV$ and  $M_\mathit{\Omega}=174.5\ GeV$ but values of  the coupling constant are different i.e. $q=10^{-7.0}$ (Fig.\ref{fig:evol_a}) and $q=10^{-6.0}$ (Fig.\ref{fig:evol_b}) respectively. Plasma temperature related to the initial points $10^{-6}$ and $10^{-2}$ of the evolution are equal to $10^7\ GeV$ and $10^4\ GeV$ respectively.}
\label{fig:Y}
\end{figure}

Now, calculating solutions of Eqs. \eqref{eq:DMFreeze7} and \eqref{eq:DMFreeze11}, under assumption of the initial condition $Y_\chi (x_0 )=Y_\mathit{\Omega}(x_0 )=10^{-18}$  and the condition \eqref{eq:DMFreeze10},  we obtain the covariant density evolution for both populations $\chi$ and $\mathit{\Omega}$. In Fig.\ref{fig:Y} we present the numerical solution for two extreme situations corresponding to different values of the coupling constant and starting point $x_0\ll 1$.

\begin{figure}[H]
\centering
\includegraphics[width=0.667\textwidth]{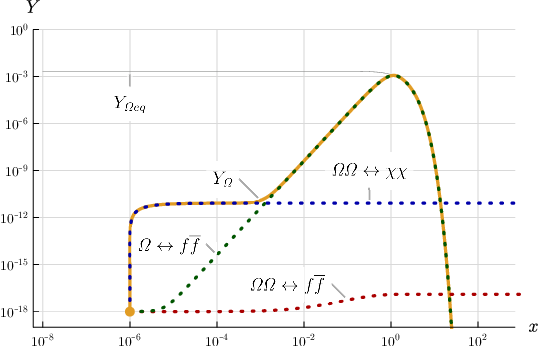}
\caption{Estimated contributions from processes $\mathit{\Omega\Omega}\leftrightarrow\chi\chi$ (blue dotted line),  $\mathit{\Omega\Omega}\leftrightarrow f\bar{f}$ (red dotted line) and  $\mathit{\Omega}\leftrightarrow f\bar{f}$ (green dotted line) to the evolution of $Y_\Omega$ (orange line) in the case described in Fig.\ref{fig:evol_a}. Take into account that single contributions does not sum in a simple way to $Y_\mathit{\Omega}$ because of nonlinearity of the Eq.\eqref{eq:Thermal4}.}
\label{fig:evol_proc}
\end{figure}

As mentioned in Sec.\ref{sec:freeze}, in the freeze-in scenario, the dark matter (DM) population begins with out-of-equilibrium particles with an extremely small number density and evolves without reaching equilibrium with the primordial plasma. In our model, the coupling constants for all essential processes are determined by a single parameter $q$ meaning the Boltzmann equations for both populations, $\chi$ and $\mathit{\Omega}$, are interconnected. We begin with a separate analysis of the processes involved in the thermal evolution of the mediator particle $\mathit{\Omega}$ (Eq.\eqref{eq:Thermal4}) using Fig.\ref{fig:evol_a} as the reference case. 
We also use Fig.\ref{fig:evol_proc} where the estimated contributions from the processes $\mathit{\Omega+\Omega}\leftrightarrow\chi+\chi$,  $\mathit{\Omega+\Omega}\leftrightarrow f+\bar{f}$ and $\mathit{\Omega}\leftrightarrow f+\bar{f}$ to the evolution of  $Y_\Omega$ are shown. 

As seen in Fig.\ref{fig:evol_a}, the thermal evolution of the dark matter covariant density $Y_\chi$, governed by the process $\chi+\chi\leftrightarrow\mathit{\Omega}+\mathit{\Omega}$, begins at $x_{0\ }={10}^{-6}$ i.e. at a temperature $T_{0}={10}^8\ GeV$ starting from an extremely small number density ${10}^{-18}$. A rapid growth of the mediator population, driven by the same creation-annihilation process, follows. As seen in Fig.\ref{fig:evol_proc}, this initially corresponds to minimal contributions from the processes $\mathit{\Omega+\Omega}\leftrightarrow f+\bar{f}$ and $\mathit{\Omega}\leftrightarrow f+\bar{f}$. Continuing analysis of  the evolution process in Fig.\ref{fig:evol_a}, we observe that with temperature decreasing, both $\chi$ and $\mathit{\Omega}$ populations enter a stability phase where their covariant densities become constant (forming a plateau in the evolution plot). At this stage, dark matter and mediator particles are in mutual equilibrium but remain out of equilibrium with the primordial SM plasma. As the temperature further decreases to $T\sim 10^5\ GeV$, we see from the Fig.\ref{fig:evol_proc} that the process $\mathit{\Omega+\Omega}\leftrightarrow f+\bar{f}$ has a negligible effect on the $\mathit{\Omega}$ density, while the decay-inverse decay process  $\mathit{\Omega}\leftrightarrow f+\bar{f}$ begins to dominate the evolution of mediator particles. Consequently, the mediator population $\mathit{\Omega}$ decouples from its equilibrium with dark matter (compare with Fig.\ref{fig:evol_proc}), while the $\chi$ particles continue their stable evolution. Next, due to interactions with SM fermions, the mediator population reaches equilibrium with the primordial plasma at a temperature of $T\sim 100\ GeV$. Finally, as the temperature drops, the $\mathit{\Omega}$ population vanishes, leaving only the frozen-in $\chi$ population with a density corresponding to the observable relic abundance. 

As we see from the Fig.\ref{fig:evol_b}, where the density evolution starts at $x_{0}={10}^{-2}$ ($T_0={10}^4\ GeV$), the $\chi$ population density rapidly increases and then reaches a plateau, while the mediator particles lose thermal equilibrium with the $\chi$ population and, as the temperature drops, eventually reaching equilibrium with the primordial SM plasma and at the temperature $\sim 100\ GeV$ begins to decrease. Therefore, regardless of the chosen starting point $x_{0}$, the evolution of both populations leads to the same final state: the $\chi$ population attains the relic abundance of DM particles, while the mediator particles decrease. However, different starting points $x_{0}$ require different values of the coupling constant $q$.

\section{Cosmological stability of dark matter} \label{Sec:stability}

One of the most peculiar attributes of the dark matter particles is their stability on cosmological time scale, larger than the age of the Universe. This means that their lifetime has to be larger than the $13.8\ Gyr\approx 4.35\cdot 10^{17}\ s$. Typically, in most dark matter models it is achieved often by assuming a discrete symmetry of the model Lagrangian. However, to obtain a complete description of the DM relic abundance, highly desirable is an explanation of DM stability from fundamental structure of the considered model. In our case stability arises in our model as the result of the spontaneous symmetry breaking under condition that dark Higgs mass, $m_\chi$, is lower than the mediator particle mass $M_\mathit{\Omega}$ i.e. $m_\chi<M_\mathit{\Omega}$ so we have deal with inverted mass hierarchy than in Higgs sector. 

To show the essence of the stability mechanism, let us recall firstly the Higgs decay into pair of neutral $Z$ bosons. Because in this case $M_Z<m_H<2M_Z$ then decay can hold via two channels  $H\rightarrow Z+Z^*$  and  $H\rightarrow Z^*+Z^*$ only.  Here,  $Z^*$  denotes a virtual off-shell mass state of the $Z$ boson. Next, on-shell and off-shell states decay into fermion--antifermion pairs, which allows preservation of the energy--momentum conservation. 

\noindent The Higgs decay into a pair $Z,Z^*$ is rather dominating i.e. its decay width is larger than the decay into a pair $Z^*,Z^*$; details of this process are given in \cite{Cahn1989TheHiggs,Grau1990Contributions,Romao1999Vector,Djouadi2008TheAnatomy,Choi2021Decays}. Consequently, the lifetime of the Higgs in the $H\rightarrow Z+Z^*$ channel is shorter than in the channel $H\rightarrow Z^*+Z^*$. Taking the above into account, let us consider the only possible dark Higgs $\chi$ decay process $\chi\rightarrow \mathit{\Omega}^*+\mathit{\Omega}^*$. Because of the energy conservation and the assumed condition  $m_\chi<M_\mathit{\Omega}$, such a process is possible only for $\mathit{\Omega}$ bosons in virtual states i.e. $\chi$ decay into virtual pair  $\mathit{\Omega}_1^*+\mathit{\Omega}_2^*$. The masses $M_1$ and $M_2$ of the virtual bosons must satisfy inequality $M_1+M_2<m_\chi$. Taking into account that the $\mathit{\Omega}_\mu$ field is coupled to neutrinos and quarks (see Eqs. \eqref{eq:Lepton5}, \eqref{eq:Quark6}), the virtual boson states decay into neutrino--antineutrino and quark--antiquark pairs. The decay width of decaying bosons is given in the Appendix \ref{Sec:dwO}. 

In Fig.\ref{fig:dwO} presented is the normalised decay width $\epsilon=\Gamma_\mathit{\Omega}/M_\mathit{\Omega}={\gamma_\mathit{\Omega}q^2}/{12\pi}$ of the process $\mathit{\Omega}\leftrightarrow f+\bar{f}$  as the function of the boson $\mathit{\Omega}$ mass.

\begin{figure}[H]
\centering
\includegraphics[width=0.667\textwidth]{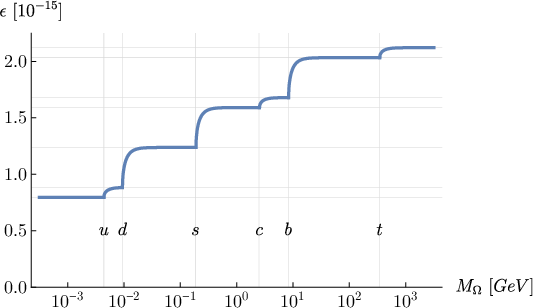}
\caption{The normalised decay width $\epsilon=\Gamma_\mathit{\Omega}/M_\mathit{\Omega}$ of the process $\mathit{\Omega}\leftrightarrow f+\bar{f}$ as function of $M_\mathit{\Omega}$ for a specific value of $q=10^{-7}$. The vertical lines indicate threshold energies for individual quark-antiquark pairs.}
\label{fig:dwO}
\end{figure}

\pagebreak
Now, in close analogy to the Higgs boson decay \cite{Cahn1989TheHiggs,Grau1990Contributions,Romao1999Vector,Djouadi2008TheAnatomy,Choi2021Decays} we can calculate the decay width of the dark particles $\chi$. The corresponding formulas take the form
\begin{subequations} \label{eq:CosmoStab1}
    \begin{align}
        \begin{split} \label{subeqn:CosmoStab1a}
            &\Gamma_{\chi\rightarrow \mathit{\Omega}^*+\mathit{\Omega}^*}=\frac{q^2 m_\chi}{32 \pi^3 \omega^2}\int_0^\omega d\mu_2 \Bigg [ \frac{\epsilon}{\left(\mu_2 -1\right)^2+\epsilon^2}\\
            &\int_0^{\left(\sqrt{\omega}-\sqrt{\mu_2}\right)^2}{d\mu_1 \frac{\epsilon}{\left(\mu_1 -1\right)^2+\epsilon^2}}\sqrt{\left(\omega-\left(\mu_1 +\mu_2\right)\right)^2-4\mu_1 \mu_2}\left(\left(\omega-\left(\mu_1 +\mu_2\right)\right)^2+8\mu_1 \mu_2\right) \Bigg ],   
        \end{split}\\
        \begin{split} \label{subeqn:CosmoStab1b}
            &\Gamma_{\chi\rightarrow \mathit{\Omega}+\mathit{\Omega}^*}=\frac{q^2 m_\chi}{32 \pi^2 \omega^2}\times\\
            &\times\int_0^{\left(\sqrt{\omega}-1\right)^2}{d\mu_1 \frac{\epsilon}{(\mu_1 -1)^2+\epsilon^2}\sqrt{\left(\omega-\left(\mu_1 +1\right)\right)^2-4\mu_1}\left(\left(\omega-\left(\mu_1 +1\right)\right)^2+8\mu_1\right)},    
        \end{split}\\
        \begin{split} \label{subeqn:CosmoStab1c}
            &\Gamma_{\chi\rightarrow \mathit{\Omega}+\mathit{\Omega}}=\frac{q^2 m_\chi}{32 \pi \omega \sqrt{\omega}}\sqrt{\omega-4}\left(\omega^2-4\omega+12\right),
        \end{split}        
    \end{align}  
\end{subequations}

\noindent where $\omega=\left(m_\chi/M_\mathit{\Omega}\right)^2$, $\epsilon=\Gamma_\mathit{\Omega}/M_\mathit{\Omega}=q^2 \gamma_\mathit{\Omega}/12 \pi$, $\mu_1=\left(M_1^*/M_\mathit{\Omega}\right)^2$, $\mu_2=\left(M_2^*/M_\mathit{\Omega}\right)^2$. 

\begin{figure}[H]
\centering
\includegraphics[width=0.667\textwidth]{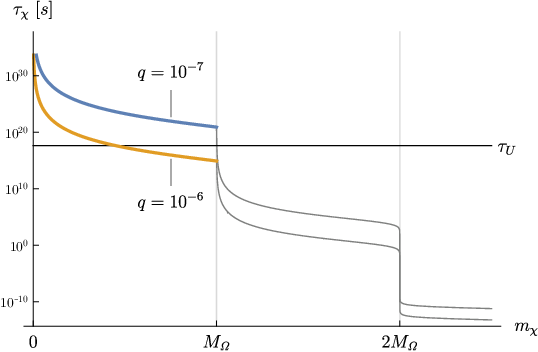}
\caption{The dark Higgs $\chi$ lifetime $\tau_\chi$ as a function of its mass $m_\chi$ given for two values of $q$: $10^{-7}$ (blue line) and $10^{-6}$ (orange line). For comparison, the age of the universe $\tau_U$ is marked with a black line. Let us remember that for dark Higgs   $m_\chi<M_\mathit{\Omega}$ (here $M_\mathit{\Omega}=174.5\ GeV$). The gray lines correspond to the case $ 2M_\mathit{\Omega}>m_\chi>M_\mathit{\Omega}$ and  $m_\chi>2M_\mathit{\Omega}$, which does not realize in our case.}
\label{fig:tau_chi}
\end{figure}

In our case, when the inequality $m_\chi<M_\mathit{\Omega}$ holds, the decay width reduces to the form  $\Gamma_{\chi\rightarrow\mathit{\Omega}^*+\mathit{\Omega}^*}$ (Eq.\eqref{subeqn:CosmoStab1a}) whereas for $2M_\mathit{\Omega}>m_\chi>M_\mathit{\Omega}$ is given by sum of $\Gamma_{\chi\rightarrow\mathit{\Omega}^*+\mathit{\Omega}^*}$ and  $\Gamma_{\chi\rightarrow\mathit{\Omega}^*+\mathit{\Omega}}$ (Eqs. \eqref{subeqn:CosmoStab1a},\eqref{subeqn:CosmoStab1b}) while for $m_\chi>2M_\mathit{\Omega}$ by $\Gamma_{\chi\rightarrow\mathit{\Omega}+\mathit{\Omega}}$. Having an appropriate decay width, we can calculate the corresponding lifetime $\tau_\chi=\hbar/\Gamma_\chi$ and check the stability condition  $\tau_\chi>\tau_U$ of dark matter. The results are demonstrated in the Fig.\ref{fig:tau_chi} for a specific $\mathit{\Omega}$ boson mass $M_\mathit{\Omega}$ and specific values of the coupling constant $q$. We see that the range of masses $m_\chi$, $M_\mathit{\Omega}$ and coupling constants $q$ satisfying the stability condition $\tau_\chi>\tau_U$ is in agreement with independently obtained  Eq.\eqref{eq:Thermal1} and the condition \eqref{eq:Thermal3}. It is worth to stress that the stability of $\chi$ particles is not achievable in two ‘standard’ sectors defined by the inequalities $2M_\mathit{\Omega}>m_\chi>M_\mathit{\Omega}$ and $m_\chi>2M_\mathit{\Omega}$.

\section{Self-interaction of the DM particles} \label{Sec:self}

An open question in astronomical observations of dark matter is the problem of its self-interaction. A widely accepted viewpoint is the so-called collisionless paradigm, supported by the analysis of weak gravitational lensing of galaxy clusters, such as the Bullet Cluster \cite{Clowe2006ADirect}. This paradigm asserts that dark matter self-interactions should be negligible to comply with constraints on structure formation after the Big Bang \cite{Randall2008Constraints,Tulin2018Dark,Harvey2015TheNongravitational}. In our model the self-interaction of $\chi$ particles is determined by the dark Higgs Lagrangian \eqref{eq:BreakDark1}, which at the tree level leads to the amplitude of the process $\chi+\chi\leftrightarrow\chi+\chi$ as given in Fig.\ref{fig:Feynman_self}. 

\begin{center}
\begin{figure}[H]
\[
\begin{aligned}
\mathbb{M}&=
\begin{tikzpicture} [baseline=(a.base)]
\begin{feynman}
\vertex[dot] (a) {};
\vertex [above left=of a] (i1) {\(k_\chi\)};
\vertex [below left=of a] (i2) {\(p_\chi\)};
\vertex [above right=of a] (f1) {\(k'_\chi\)};
\vertex [below right=of a] (f2) {\(p'_\chi\)};
\diagram* [horizontal=i1 to f1] { 
  (i1) -- [draw=none] (i2),
  (i1) -- [scalar] (a),
  (i2) -- [scalar] (a),
  (a)-- [scalar] (f1), 
  (a)-- [scalar] (f2), 
}; 
\vertex [right=1 of a] {\(3iq^2\frac{m_\chi^2}{M_\mathit{\Omega}^2}\)};
\end{feynman}
\end{tikzpicture}
+
\begin{tikzpicture} [baseline=(a.base)]
\begin{feynman}
\vertex[dot] (a) {};
\vertex [above left=of a] (i1) {\(k_\chi\)};
\vertex [below left=of a] (i2) {\(p_\chi\)};
\vertex[dot] [right=of a] (b) {};
\vertex [above right=of b] (f1) {\(k'_\chi\)};
\vertex [below right=of b] (f2) {\(p'_\chi\)};
\diagram* [vertical'=a to b,horizontal=i1 to a] { 
  (i1) -- [draw=none] (i2),
  (i1) -- [scalar] (a),
  (i2) -- [scalar] (a),
  (a) -- [scalar] (b),
  (b)-- [scalar] (f1), 
  (b)-- [scalar] (f2), 
}; 
\vertex [left=1 of a] {\(3iq\frac{m_\chi^2}{M_\mathit{\Omega}}\)};
\vertex [right=1 of b] {\(3iq\frac{m_\chi^2}{M_\mathit{\Omega}}\)};
\vertex [right=0.75 of a] (c);
\vertex [above=0.25 of c] {\(\frac{i}{\left(k_\chi+p_\chi\right)^2-m_\chi^2}\)};
\end{feynman}
\end{tikzpicture}
+ \\
&+\begin{tikzpicture} [baseline=(center.base)]
\begin{feynman}
\vertex at (0,0) (center);
\vertex at (0,1) (i1) {\(k_\chi\)};
\vertex at (0,-1) (i2) {\(p_\chi\)};
\vertex[dot] [right=of i1] (a) {};
\vertex[dot] [right=of i2] (b) {};
\vertex [right=of a] (f1) {\(k'_\chi\)};
\vertex [right=of b] (f2) {\(p'_\chi\)};
\diagram* [vertical'=a to b,horizontal=i1 to a] { 
  (i1) -- [scalar] (a),
  (i2) -- [scalar] (b),
  (a) -- [scalar, edge label=\(\frac{i}{l^2-m_\chi^2}\)] (b),
  (a)-- [scalar] (f1), 
  (b)-- [scalar] (f2), 
}; 
\vertex [above=0.5 of a] {\(3iq\frac{m_\chi^2}{M_\mathit{\Omega}}\)};
\vertex [below=0.5 of b] {\(3iq\frac{m_\chi^2}{M_\mathit{\Omega}}\)};
\end{feynman}
\end{tikzpicture}
+
\begin{tikzpicture} [baseline=(center.base)]
\begin{feynman}
\vertex at (0,0) (center);
\vertex at (0,1) (i1) {\(k_\chi\)};
\vertex at (0,-1) (i2) {\(p_\chi\)};
\vertex[dot] [right=of i1] (a) {};
\vertex[dot] [right=of i2] (b) {};
\vertex [right=of a] (f1) {\(k'_\chi\)};
\vertex [right=of b] (f2) {\(p'_\chi\)};
\diagram* [vertical'=a to b,horizontal=i1 to a] { 
  (i1) -- [scalar] (a),
  (i2) -- [scalar] (b),
  (a) -- [scalar] (b),
  (a)-- [scalar] (f2), 
  (b)-- [scalar] (f1), 
}; 
\vertex [above=0.5 of a] {\(3iq\frac{m_\chi^2}{M_\mathit{\Omega}}\)};
\vertex [below=0.5 of b] {\(3iq\frac{m_\chi^2}{M_\mathit{\Omega}}\)};
\vertex [below=1 of a] (c);
\vertex [left=0 of c] {\(\frac{i}{r^2-m_\chi^2}\)};
\end{feynman}
\end{tikzpicture}
\end{aligned}
\]
where $l=k'_\chi-k_\chi$, $r=p'_\chi-k_\chi$
\caption[]{Amplitude of the self-interaction process for the $\chi$ field.}
\label{fig:Feynman_self}
\end{figure}
\end{center}

\begin{figure}[H]
    \centering
    \includegraphics[width=0.667\textwidth]{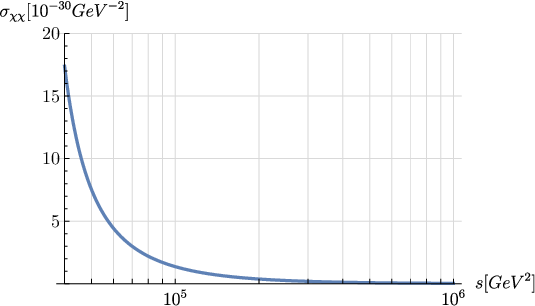}
    \caption[]{The cross section of the self-interaction process $\chi+\chi\leftrightarrow\chi+\chi$ given for $q=10^{-6}$, $m_\chi=100\ GeV$ and $M_\mathit{\Omega}=174.5\ GeV$.}
    \label{fig:sigmachi}
\end{figure}

The corresponding cross section $\sigma_{\chi\chi}(s)$ is provided in Appendix \ref{Sec:csxx}. For specific values of mass and coupling constants, $\sigma_{\chi\chi}(s)$ is illustrated in Fig.\ref{fig:sigmachi}.

The experimental upper bound for the self-interaction of dark matter particles is expressed by the inequality \cite{Navas2024Review,Randall2008Constraints,Tulin2018Dark,Harvey2015TheNongravitational}
\begin{equation} \label{eq:Self1}
    \frac{\sigma_{\chi\chi}}{m_\chi}<0.47\frac{cm^2}{g}=\frac{2.145\cdot 10^3}{GeV^3}.
\end{equation}
Using Eq.\eqref{eq:Thermal1} and the explicit form of $\sigma_{\chi\chi}(s)$ we arive at 
\begin{equation} \label{eq:Self2}
    \frac{\sigma_{\chi\chi}}{m_\chi}\leq \frac{81q^4m_\chi}{16\pi M_\mathit{\Omega}^3}<\frac{q^4}{M_\mathit{\Omega}^3}.
\end{equation}
Considering the limitation \eqref{eq:Thermal1}, we conclude that our model complies with the empirical constraint \eqref{eq:Self1}.

\section{Resume and conclusions} \label{Sec:resume}

In this paper, we propose a simple yet nontrivial extension of the Standard Model of weak interaction, based on the $SU(2)_L\times U(1)_Y\times U(1)_C$ electroweak group. Our motivation arises from identifying the globally conserved charge $\Omega=\mathrm{B}-\mathrm{L}-\mathrm{Q}$  as a neutrino charge. Our goal was to find a neutrino counterpart to the electric current. To achieve this, we require that a gauge field, $\mathit{\Omega}_\mu$, in the lepton sector interacts exclusively with the neutrino. This requirement imposes certain conditions on the connection coefficients and the generators (charges) of the $U(1)_C$ group in the lepton sector (Eqs. \eqref{eq:Lepton2}, \eqref{eq:Lepton3}, \eqref{eq:Lepton4}). The consistency of the model enforces a specific form of connection and charges of the $U(1)_C$ in the quark sector as well (Eqs. \eqref{eq:Quark2}, \eqref{eq:Quark3}, \eqref{eq:Quark4}, \eqref{eq:Quark5}). This allows us to identify both the lepton and quark neutrino currents, along with their corresponding Lagrangians (Eqs. \eqref{eq:Lepton5}, \eqref{eq:Lepton6}, \eqref{eq:Lepton7} and Eqs. \eqref{eq:Quark6}, \eqref{eq:Quark7}, \eqref{eq:Quark8}). The gauge fields Lagrangian is provided in Eqs. \eqref{eq:GaugeL1} and \eqref{eq:GaugeL2}. 

To generate masses for gauge bosons and fermions, we perform a two-stage process of spontaneous symmetry breaking. In the first stage, described in  Sec.\ref{Sec:breakingHiggs}, we follow the SM Higgs mechanism, breaking the $SU(2)_L\times U(1)_Y\times U(1)_C$ down to $U(1)_Q\times U(1)_\Omega$ using the Goldstone--Higgs iso-dublet. This results in mass generation for the $W_\mu^\pm$ and $Z_\mu$ gauge bosons, as well as Dirac masses for fundamental fermions, while  $A_\mu$  and   $\mathit{\Omega}_\mu$  remain massless. The final form of the C–charges is given in Eqs. \eqref{eq:BreakHiggs6} and \eqref{eq:BreakHiggs7}), while a modified relationship between the $W_\mu^\pm$ and $Z_\mu$  masses is presented in Eq.\eqref{eq:BreakHiggs10}.

To avoid problems associated with the long-range $\mathit{\Omega}_\mu$, we opt for a spontaneously broken $U(1)$ symmetry scenario, using an additional Higgs iso-scalar $\chi$. 
Because the field $\chi$ interacts only with themselves and the mediator field $\mathit{\Omega}_\mu$ (see Eq.\eqref{eq:BreakDark1}), then we identify the particle population $\chi$ as scalar dark matter. Using the Boltzmann evolution equations (Eqs. \eqref{eq:DMFreeze11}, \eqref{eq:DMFreeze12}) for particle number densities of both the $\chi$ and the $\mathit{\Omega}_\mu$ populations, we determine the relic abundance of $\chi$, selecting masses and the coupling constant to match the observed dark matter relic abundance. Under this condition, we find a relationship between the masses and the coupling constant, determining this coupling constant to be in the range $q\sim 10^{-8.5} g$  to $\sim 10^{-6} g$. Cosmological stability of $\chi$ particles is achieved as a consequence of the inverted spectral condition, $m_\chi<M_\mathit{\Omega}$, between the dark Higgs $\chi$ and the gauge boson $\mathit{\Omega}_\mu$ masses (Sec.\ref{Sec:stability}).
\begin{figure}[H]
    \centering
    \includegraphics[width=0.667\textwidth]{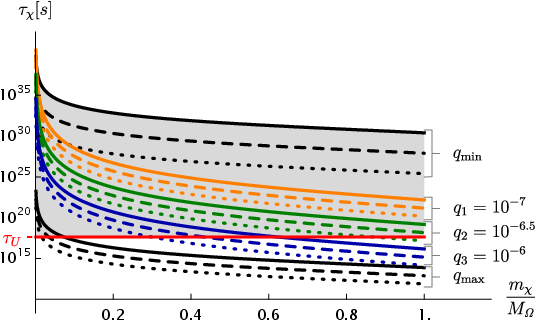}
    $$\scriptstyle q_{\mathrm{min}}=2.4\cdot 10^{-6}\omega^{-\frac14}, \quad q_{\mathrm{max}}=2.4\cdot 10^{-9}\left(\frac{m_\chi}{\omega}\right)^\frac14$$
    \caption[]{The $\chi$ particle lifetime $\tau_\chi$ as a function of the mass ratio $m\chi/M_\mathit{\Omega}$ for three values of $m_\chi$: $10\ GeV$ (solid lines), $100\ GeV$ (dashed lines), $1\ TeV$ (dotted lines) and for various coupling constant $q$ values: minimal and maximal ones ($m_\chi$ and $M_\mathit{\Omega}$ dependent) as well as for three intermediate constant values: $10^{-7}$, $10^{-6.5}$, $10^{-6}$. The admissible region of parameters $m_\chi$, $M_\mathit{\Omega}$ and $q$, bounded by the age of the Universe $\tau_U$ and $\tau_\chi (q_{\mathrm{min}}(m_\chi/M_\mathit{\Omega}))$, is shaded.}
    \label{fig:tau_chi_2}
\end{figure}

Now, assuming the age of the Universe as a symbolic lower bound of the $\chi$ particles' lifetime (see the red line in Fig.\ref{fig:tau_chi_2}) and taking into account the limits \eqref{eq:Thermal3} on $q$ enforced by the model, we can identify the region of admissible values of the masses $m_\chi$ and $M_\mathit{\Omega}$  as well as of the coupling constant $q$ (see Fig.\ref{fig:tau_chi_2}).Notice that $\chi$ particles are not absolutely stable because of an upper limit for their lifetime $\tau_\chi$  (upper black solid line in Fig.\ref{fig:tau_chi_2}). However, they are cosmologically stable i.e. they are stable in the cosmological scale.

Summarizing, our minimal extension of the SM introduces an additional gauge field  $\mathit{\Omega}_\mu$ (the mediator field) and a scalar dark Higgs field $\chi$, while naturally incorporating the right-chiral neutrino. Furthermore, the model exhibits a feeble interaction with a very small coupling constant $q$ (see Eq.\eqref{eq:Thermal3}), characteristic of freeze-in processes. As noted in Sec.\ref{Sec:lepton}, this explains the experimental non-observability of the right neutrino, due to decoupling caused by the very small $q$. In this extension, neutrinoless double $\beta$ decay is forbidden because of the conservation of the global neutrino charge. The Boltzmann evolution equations (Eqs. \eqref{eq:DMFreeze7}, \eqref{eq:Thermal4}) for the particle number densities of both $\chi$ and $\mathit{\Omega}_\mu$ populations lead to the correct dark matter relic abundance. They also explain the close relationship between the number density evolution of dark Higgs $\chi$ and mediator $\mathit{\Omega}_\mu$ populations (see for example, Fig.\ref{fig:Y}). An intriguing feature of the model is its intrinsic discrete structural symmetry, linking charged leptons with neutrinos, and up with down quarks. 

\begin{figure}[H]
\centering
\includegraphics[width=0.667\textwidth]{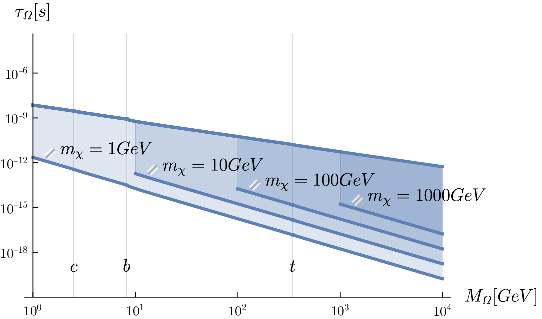}
\caption{The $\mathit{\Omega}$ boson lifetime as a function of its mass $M_\mathit{\Omega}$ in the area admissible by the condition \eqref{eq:Thermal3} for various choices of $\chi$ particle mass $m_\chi$.}
\label{fig:tau}
\end{figure}

In Fig.\ref{fig:tau} we show the admissible by the inequality \eqref{eq:Thermal3} region of the lifetime of boson $\mathit{\Omega}$ as its mass function.

A primary challenge lies in the experimental determination of the constants $g_V^\nu$, $g_V^l$, $g_A^l$, $g_V^u$, $g_V^d$ in Eqs. \eqref{eq:Lepton7} and \eqref{eq:Quark8} with sufficient precision, given the small value of the coupling  $q=g\cos\theta\sin\varphi$. These constants describe the coupling of lepton and quark currents with the neutral $Z$ boson. Deviation from theirs standard form is of the order $q^2$ so it is very small. Furthermore, a direct identification of the dark Higgs boson is achievable only through the process $\mathit{\Omega}+\mathit{\Omega}\rightarrow\chi+\chi$ or $\mathit{\Omega}+\mathit{\Omega}\rightarrow\chi$. This limitation can be a cause of the current non-observability of dark matter particles in experimental settings. Thus, it seems that identification of the mediator boson $\mathit{\Omega}_\mu$ is  crucial to the identification of the dark Higgs. An experimental method for detecting this boson, produced via $f+\bar{f}\rightarrow\mathit{\Omega}$ and undergoing subsequent decay $\mathit{\Omega}\rightarrow f'+\bar{f'}$ (refer to Fig. \ref{fig:tau}), appears feasible, given its relatively long lifetime. 
\clearpage

\appendix
\section{Gauge field lagrangian} \label{Sec:gaugeLAp}

The gauge part of the full Lagrangian takes the manifestly covariant form

\begin{equation} \label{eq:A1}
\begin{split}
\mathcal{L}_{\mathrm{gauge}} &= -\frac{1}{2}\left(\partial_{\mu }W_{\nu }^{+}-\partial_{\nu }W_{\mu }^{+}\right)\left(\partial^{\mu }W^{-\nu }-\partial^{\nu }W^{-\mu }\right)-\frac{1}{4}F_{\mu \nu }F^{\mu \nu }-\frac{1}{4}Z_{\mu \nu }Z^{\mu \nu }-\frac{1}{4}\mathit{\Omega}_{\mu \nu }\mathit{\Omega}^{\mu \nu }+\\
&-\frac{ig}{2}\Bigg\{\Big[\Big(\sin \theta\left(A_{\mu }W_{\nu }^{-}-A_{\nu }W_{\mu }^{-}\right)+\cos \theta \cos \varphi\left(Z_{\mu }W_{\nu }^{-}-Z_{\nu }W_{\mu }^{-}\right)+
\\
&-\cos \theta \sin \varphi\left(\mathit{\mathit{\Omega}}_\mu W_{\nu }^{-}-\mathit{\Omega}_{\nu }W_{\mu }^{-}\right)\Big)\left(\partial^{\mu }W^{+\nu }-\partial^{\nu }W^{+\mu }\right)-\mathrm{h.c.}\Big]+
\\
&-\left(\sin \theta F_{\mu \nu }+\cos \theta \cos \varphi Z_{\mu \nu }- \cos \theta \sin \varphi \mathit{\Omega}_{\mu \nu }\right)
\left(W^{+\mu }W^{-\nu }-W^{+\nu }W^{-\mu }\right)\Bigg\}+
\\
&-\frac{g^{2}}{2}\Bigg\{\Big[\sin \theta\left(A_{\mu }W_{\nu }^{-}-A_{\nu }W_{\mu }^{-}\right)+\cos \theta \cos \varphi\left(Z_{\mu }W_{\nu }^{-}-Z_{\nu }W_{\mu }^{-}\right)+
\\
&-\cos \theta \sin \varphi\left(\mathit{\mathit{\Omega}}_\mu W_{\nu }^{-}-\mathit{\Omega}_{\nu }W_{\mu }^{-}\right)\Big]
\Big(\sin \theta\left(A^{\mu }W^{+\nu }-A^{\nu }W^{+\mu }\right)+
\\
&+\cos \theta \cos \varphi\left(Z^{\mu }W^{+\nu }-Z^{\nu }W^{+\mu }\right)
-\cos \theta \sin \varphi\left(\mathit{\Omega}^{\mu }W^{+\nu }-\mathit{\Omega}^{\nu }W^{+\mu }\right)\Big)+
\\
&-\left(W_{\mu }^{+}W_{\nu }^{-}-W_{\nu }^{+}W_{\mu }^{-}\right)\left(W^{+\mu }W^{-\nu }-W^{+\nu }W^{-\mu }\right)\Bigg\}.
\end{split}
\end{equation}

As before, in the limit $ \varphi \rightarrow 0 $ and $ M_\mathit{\Omega}\rightarrow 0 $, $ \mathcal{L}_{\mathrm{gauge}}\rightarrow\mathcal{L}_{\mathrm{gaugeSM}} $.

\section{Relationship between the \texorpdfstring{$C$}--charges and \texorpdfstring{$z$}--charges}

In the Sec.\ref{Sec:gaugeL}  we give also a proof of gauge anomalies cancellation by using the results of the Refs. \cite{Applequist2003Nonexotic} and \cite{Ekstedt2018Minimal}. Here, for reader convenience,  we present  the relationship between labeling of $C$-charges in our paper and $z$-charges in the Ref.\cite{Ekstedt2018Minimal}. It is the following:

\begin{equation} \label{eq:B1}
z_l\equiv c_1,\quad z_k\equiv c_2,\quad z_e\equiv c_3,\quad z_q\equiv \tilde{c}_1,\quad z_u\equiv \tilde{c}_2,\quad z_d\equiv \tilde{c}_3,\quad z_H\equiv c_H.    
\end{equation}

\section{Cross section of the process \texorpdfstring{$\mathit{\Omega\Omega}\leftrightarrow\chi\chi$}{OO<->xx}} \label{Sec:csOOxx}

The cross section $\sigma\left(q,s,m_\chi,M_\mathit{\Omega}\right)$ of the process $\mathit{\Omega+\Omega}\leftrightarrow\chi+\chi$ arising from its amplitude from  Fig.\ref{fig:FeynmanOOxx}, as a function of the Mandelstam variable $s$ is given by the fallowing formula:

\begin{equation} \label{eq:A2}
    \begin{aligned}
        \sigma_{\mathit{\Omega\Omega}\leftrightarrow\chi\chi}(s)&=
        \frac{q^4 \sqrt{s-4 M_\mathit{\Omega}^2}}{16 \pi  M_\mathit{\Omega}^4 s \sqrt{s-4 m_\chi^2}}\Bigg[
        \frac{1}{\left(m_\chi^2-s\right)^2}
        \Big( 4 m_\chi^8-4 m_\chi^6 \left(8 M_\mathit{\Omega}^2+5 s\right)+\\
        &+m_\chi^4 \left(192 M_\mathit{\Omega}^4+16 M_\mathit{\Omega}^2 s+25 s^2\right)-4 m_\chi^2 M_\mathit{\Omega}^2 s \left(24 M_\mathit{\Omega}^2+5 s\right)+12 M_\mathit{\Omega}^4 s^2\Big)+\\
        &+\frac{2 \left(m_\chi^8-8 m_\chi^6 M_\mathit{\Omega}^2+24 m_\chi^4 M_\mathit{\Omega}^4-32 m_\chi^2 M_\mathit{\Omega}^6+48 M_\mathit{\Omega}^8\right)}{m_\chi^4-4 m_\chi^2 M_\mathit{\Omega}^2+M_\mathit{\Omega}^2 s}+\\
        &-\frac{4}{\sqrt{s-4 m_\chi^2}\sqrt{s-4 M_\mathit{\Omega}^2} \left(2 m_\chi^4-3 m_\chi^2 s+s^2\right) } \Big(3 m_\chi^{10}-m_\chi^8 \left(24 M_\mathit{\Omega}^2+11 s\right)+\\
        &+m_\chi^6 \left(56 M_\mathit{\Omega}^4+52 M_\mathit{\Omega}^2 s+5 s^2\right)-2 m_\chi^4 \left(80 M_\mathit{\Omega}^6+4 M_\mathit{\Omega}^4 s+19 M_\mathit{\Omega}^2 s^2\right)+\\
        &+2 m_\chi^2 M_\mathit{\Omega}^2 \left(-24 M_\mathit{\Omega}^6+56 M_\mathit{\Omega}^4 s-6 M_\mathit{\Omega}^2 s^2+5 s^3\right)+24 M_\mathit{\Omega}^6 s \left(2 M_\mathit{\Omega}^2-s\right)\Big)\times\\
        &\times\log \left(\frac{-\sqrt{s-4 m_\chi^2} \sqrt{s-4 M_\mathit{\Omega}^2}+2 m_\chi^2-s}{\sqrt{s-4 m_\chi^2}\sqrt{s-4 M_\mathit{\Omega}^2}+2 m_\chi^2-s}\right)
        +8 M_\mathit{\Omega}^2 s
        \Bigg].  
    \end{aligned}
\end{equation}

\section{Cross section of the process \texorpdfstring{$\mathit{\Omega\Omega}\leftrightarrow f\bar{f}$}{OO<->ff}} \label{Sec:csOOff}

The cross section $\sigma_{\mathit{\Omega\Omega}\leftrightarrow f\bar{f}} (s)$ of the process $\mathit{\Omega+\Omega}\leftrightarrow f+\bar{f}$, calculated under assumption of small fermionic masses in comparison with $M_\mathit{\Omega}$, is given in the following formula:

\begin{equation} \label{eq:A3}
    \sigma_{\mathit{\Omega\Omega}\leftrightarrow f\bar{f}}(s)=\frac{131 q^4}{108\pi s^2}\left(\frac{s^2+4M_\mathit{\Omega}^4}{s-2M_\mathit{\Omega}^2}\log{\frac{1+\sqrt{1-\dfrac{4M_\mathit{\Omega}^2}{s}}}{1-\sqrt{1-\dfrac{4M_\mathit{\Omega}^2}{s}}}}-s\sqrt{1-\frac{4M_\mathit{\Omega}^2}{s}}\right),
\end{equation}

while its plot is presented in the Fig.\ref{fig:csOOff}.

\section{Decay width of the process \texorpdfstring{$\mathit{\Omega}\leftrightarrow f\bar{f}$}{O<->ff}} \label{Sec:dwO}

The decay width $\Gamma$ of the process $\mathit{\mathit{\Omega}}\leftrightarrow f+\bar{f}$ can be easily calculated and is given by the formula

\begin{equation} \label{eq:A4}
    \begin{split}
        \Gamma_\Omega&=\frac{q^2M_\mathit{\Omega}}{12\pi}\sum_f{\Omega_{f}^2n_{f}\sqrt{1-\frac{4m_{f}^2}{M_\mathit{\Omega}^2}}\left(1+\frac{2m_{f}^2}{M_\mathit{\Omega}^2}\right)}\equiv\frac{q^2M_\mathit{\Omega}}{12\pi}\gamma_\mathit{\Omega},  
    \end{split}
\end{equation}
where sum is definied over all fundamental fermions $f=\nu_e,\nu_\mu,\nu_\tau,e,\mu,\tau,u,c,t,d,s,b$;  taking into account their flavour and colour. Fermion masses satisfy inequality $M_\mathit{\Omega}>2m_f$. $m_{f}$ is the mass of fermion $f_i$ (neutrino mass $m_\nu\approx 0$), while $\Omega_{f}$ is its $\Omega$-charge and $n_{f}$ denotes number of colours for quarks ($3$ for each flavour) while for leptons is equal to $1$. Taking into account that $\Omega_\nu=-1,\Omega_e=\Omega_\mu=\Omega_\tau=0,\Omega_u=\Omega_c=\Omega_t=-\frac13,\Omega_d=\Omega_s=\Omega_b=+\frac23$, we obtain that

\begin{equation}
\begin{split}
    \gamma_\mathit{\Omega}&=
    3+\frac{4}{3}\sum_{f=d,s,b}{\sqrt{1-\frac{4m_{f}^2}{M_\mathit{\Omega}^2}}\left(1+\frac{2m_{f}^2}{M_\mathit{\Omega}^2}\right)\theta\left(M_\mathit{\Omega}-2m_f\right)}+\\
    &+\frac{1}{3}\sum_{f=u,c,t}{\sqrt{1-\frac{4m_{f}^2}{M_\mathit{\Omega}^2}}\left(1+\frac{2m_{f}^2}{M_\mathit{\Omega}^2}\right)\theta\left(M_\mathit{\Omega}-2m_f\right)},
    \end{split}
\end{equation}
\noindent where $\theta\left(M_\mathit{\Omega}-2m_f\right)$ denotes the Heaviside \emph{theta} function.

Notice, that no contributions of the charged leptons to $\Gamma_\mathit{\Omega}$ width because theirs $\Omega$ charge is equal to zero. The formula \eqref{eq:A4} can be compared with decay width of the $Z$ boson into fermion-antifermion pair \cite{Belanger2020Dark}. In calculations of the normalized width, it is enough to use the quark pole masses instead of the running quark masses.

\section{\texorpdfstring{$\chi$}{x} particles self-interaction cross section} \label{Sec:csxx}

The $\chi$ particles self-interaction cross section  $\sigma_{\chi\chi} (s)$, calculated with use  of the amplitude presented in the Fig.\ref{fig:Feynman_self} is given by the following formula:

\begin{equation} \label{eq:A5}
\begin{split}
    \sigma_{\chi\chi}(s)&=\frac{9}{16\pi}\left(\frac{qm_\chi}{M_\mathit{\Omega}}\right)^4\Bigg [
    \frac{\left( 2m_\chi^2-s\right)\left( 4m_\chi^2-s\right)\left( 5m_\chi^2+s\right)\left( 6m_\chi^4-2m_\chi^2s-s^2\right)}{s\left( m_\chi^2-s\right)^2\left(4m_\chi^2-s\right)\left( 6m_\chi^4-5m_\chi^2s+s^2\right)}+\\
    &+\frac{12m_\chi^2\left(m_\chi^2-s\right)\left(3m_\chi^2-s\right)\left( 5m_\chi^4-3m_\chi^2s+s^2\right)\log{\frac{m_\chi^2}{s-3m_\chi^2}}}{s\left( m_\chi^2-s\right)^2\left(4m_\chi^2-s\right)\left( 6m_\chi^4-5m_\chi^2s+s^2\right)} \Bigg ].
\end{split}
\end{equation}

\begin{acknowledgments}
We wish to thank to Karol Ławniczak for his help  with numerical calculations and for discussions.
We also gratefully acknowledge Polish high-performance computing infra\-structure PLGrid (HPC Center: ACK Cyfronet AGH) for providing computer facilities and support within computational grant no. PLG/2023/016792.
\end{acknowledgments}


\begin{thebibliography}{37}%
\makeatletter
\providecommand \@ifxundefined [1]{%
 \@ifx{#1\undefined}
}%
\providecommand \@ifnum [1]{%
 \ifnum #1\expandafter \@firstoftwo
 \else \expandafter \@secondoftwo
 \fi
}%
\providecommand \@ifx [1]{%
 \ifx #1\expandafter \@firstoftwo
 \else \expandafter \@secondoftwo
 \fi
}%
\providecommand \natexlab [1]{#1}%
\providecommand \enquote  [1]{``#1''}%
\providecommand \bibnamefont  [1]{#1}%
\providecommand \bibfnamefont [1]{#1}%
\providecommand \citenamefont [1]{#1}%
\providecommand \@href[1]{\@@startlink{#1}\@@href}%
\providecommand \@@href[1]{\endgroup#1\@@endlink}%
\providecommand \@sanitize@url [0]{\catcode `\\12\catcode `\$12\catcode `\&12\catcode `\#12\catcode `\^12\catcode `\_12\catcode `\%12\relax}%
\providecommand \@@startlink[1]{}%
\providecommand \@@endlink[0]{}%
\providecommand \@url [1]{\endgroup\@href {#1}{\urlprefix }}%
\providecommand \urlprefix  [0]{URL }%
\providecommand \Eprint [0]{\href }%
\providecommand \doibase [0]{https://doi.org/}%
\providecommand \selectlanguage [0]{\@gobble}%
\providecommand \bibinfo  [0]{\@secondoftwo}%
\providecommand \bibfield  [0]{\@secondoftwo}%
\providecommand \translation [1]{[#1]}%
\providecommand \BibitemOpen [0]{}%
\providecommand \bibitemStop [0]{}%
\providecommand \bibitemNoStop [0]{.\EOS\space}%
\providecommand \EOS [0]{\spacefactor3000\relax}%
\providecommand \BibitemShut  [1]{\csname bibitem#1\endcsname}%
\let\auto@bib@innerbib\@empty
\bibitem [{\citenamefont {Appelquist}\ \emph {et~al.}(2003)\citenamefont {Appelquist}, \citenamefont {Dobrescu},\ and\ \citenamefont {Hopper}}]{Applequist2003Nonexotic}%
  \BibitemOpen
  \bibfield  {author} {\bibinfo {author} {\bibfnamefont {T.}~\bibnamefont {Appelquist}}, \bibinfo {author} {\bibfnamefont {B.~A.}\ \bibnamefont {Dobrescu}},\ and\ \bibinfo {author} {\bibfnamefont {A.~R.}\ \bibnamefont {Hopper}},\ }\bibfield  {title} {\bibinfo {title} {{Nonexotic neutral gauge bosons}},\ }\href {https://doi.org/10.1103/PhysRevD.68.035012} {\bibfield  {journal} {\bibinfo  {journal} {Physical Review D}\ }\textbf {\bibinfo {volume} {68}},\ \bibinfo {pages} {035012} (\bibinfo {year} {2003})},\ \Eprint {https://arxiv.org/abs/hep-ph/0212073} {arXiv:hep-ph/0212073} \BibitemShut {NoStop}%
\bibitem [{\citenamefont {Langacker}(2009)}]{Langacker2009ThePhysics}%
  \BibitemOpen
  \bibfield  {author} {\bibinfo {author} {\bibfnamefont {P.}~\bibnamefont {Langacker}},\ }\bibfield  {title} {\bibinfo {title} {{The Physics of Heavy Z0 Gauge Bosons}},\ }\href {https://doi.org/10.1103/RevModPhys.81.1199} {\bibfield  {journal} {\bibinfo  {journal} {Reviews of Modern Physics}\ }\textbf {\bibinfo {volume} {81}},\ \bibinfo {pages} {1199} (\bibinfo {year} {2009})},\ \Eprint {https://arxiv.org/abs/0801.1345} {arXiv:0801.1345} \BibitemShut {NoStop}%
\bibitem [{\citenamefont {Ekstedt}\ \emph {et~al.}(2018)\citenamefont {Ekstedt}, \citenamefont {Enberg}, \citenamefont {Ingelman}, \citenamefont {L{\"{o}}fgren},\ and\ \citenamefont {Mandal}}]{Ekstedt2018Minimal}%
  \BibitemOpen
  \bibfield  {author} {\bibinfo {author} {\bibfnamefont {A.}~\bibnamefont {Ekstedt}}, \bibinfo {author} {\bibfnamefont {R.}~\bibnamefont {Enberg}}, \bibinfo {author} {\bibfnamefont {G.}~\bibnamefont {Ingelman}}, \bibinfo {author} {\bibfnamefont {J.}~\bibnamefont {L{\"{o}}fgren}},\ and\ \bibinfo {author} {\bibfnamefont {T.}~\bibnamefont {Mandal}},\ }\bibfield  {title} {\bibinfo {title} {{Minimal anomalous U(1) theories and collider phenomenology}},\ }\href {https://doi.org/10.1007/JHEP02(2018)152} {\bibfield  {journal} {\bibinfo  {journal} {Journal of High Energy Physics}\ }\textbf {\bibinfo {volume} {2018}},\ \bibinfo {pages} {152} (\bibinfo {year} {2018})},\ \Eprint {https://arxiv.org/abs/1712.03410} {arXiv:1712.03410} \BibitemShut {NoStop}%
\bibitem [{\citenamefont {Arcadi}\ \emph {et~al.}(2018)\citenamefont {Arcadi}, \citenamefont {Dutra}, \citenamefont {Ghosh}, \citenamefont {Lindner}, \citenamefont {Mambrini}, \citenamefont {Pierre}, \citenamefont {Profumo},\ and\ \citenamefont {Queiroz}}]{Arcadi2018TheWaning}%
  \BibitemOpen
  \bibfield  {author} {\bibinfo {author} {\bibfnamefont {G.}~\bibnamefont {Arcadi}}, \bibinfo {author} {\bibfnamefont {M.}~\bibnamefont {Dutra}}, \bibinfo {author} {\bibfnamefont {P.}~\bibnamefont {Ghosh}}, \bibinfo {author} {\bibfnamefont {M.}~\bibnamefont {Lindner}}, \bibinfo {author} {\bibfnamefont {Y.}~\bibnamefont {Mambrini}}, \bibinfo {author} {\bibfnamefont {M.}~\bibnamefont {Pierre}}, \bibinfo {author} {\bibfnamefont {S.}~\bibnamefont {Profumo}},\ and\ \bibinfo {author} {\bibfnamefont {F.~S.}\ \bibnamefont {Queiroz}},\ }\bibfield  {title} {\bibinfo {title} {{The waning of the WIMP? A review of models, searches, and constraints}},\ }\href {https://doi.org/10.1140/epjc/s10052-018-5662-y} {\bibfield  {journal} {\bibinfo  {journal} {The European Physical Journal C}\ }\textbf {\bibinfo {volume} {78}},\ \bibinfo {pages} {203} (\bibinfo {year} {2018})},\ \Eprint {https://arxiv.org/abs/1703.07364} {arXiv:1703.07364} \BibitemShut {NoStop}%
\bibitem [{\citenamefont {K{\"{o}}rs}\ and\ \citenamefont {Nath}(2004)}]{Kors2004AStuckelberg}%
  \BibitemOpen
  \bibfield  {author} {\bibinfo {author} {\bibfnamefont {B.}~\bibnamefont {K{\"{o}}rs}}\ and\ \bibinfo {author} {\bibfnamefont {P.}~\bibnamefont {Nath}},\ }\bibfield  {title} {\bibinfo {title} {{A Stueckelberg extension of the Standard Model}},\ }\href {https://doi.org/10.1016/j.physletb.2004.02.051} {\bibfield  {journal} {\bibinfo  {journal} {Physics Letters B}\ }\textbf {\bibinfo {volume} {586}},\ \bibinfo {pages} {366} (\bibinfo {year} {2004})},\ \Eprint {https://arxiv.org/abs/hep-ph/0402047} {arXiv:hep-ph/0402047} \BibitemShut {NoStop}%
\bibitem [{\citenamefont {Burgess}\ \emph {et~al.}(2001)\citenamefont {Burgess}, \citenamefont {Pospelov},\ and\ \citenamefont {ter Veldhuis}}]{Burgess2001TheMinimal}%
  \BibitemOpen
  \bibfield  {author} {\bibinfo {author} {\bibfnamefont {C.}~\bibnamefont {Burgess}}, \bibinfo {author} {\bibfnamefont {M.}~\bibnamefont {Pospelov}},\ and\ \bibinfo {author} {\bibfnamefont {T.}~\bibnamefont {ter Veldhuis}},\ }\bibfield  {title} {\bibinfo {title} {{The Minimal Model of nonbaryonic dark matter: a singlet scalar}},\ }\href {https://doi.org/10.1016/S0550-3213(01)00513-2} {\bibfield  {journal} {\bibinfo  {journal} {Nuclear Physics B}\ }\textbf {\bibinfo {volume} {619}},\ \bibinfo {pages} {709} (\bibinfo {year} {2001})},\ \Eprint {https://arxiv.org/abs/hep-ph/0011335} {arXiv:hep-ph/0011335} \BibitemShut {NoStop}%
\bibitem [{\citenamefont {Dobrescu}(2005)}]{Dobrescu2005Masslss}%
  \BibitemOpen
  \bibfield  {author} {\bibinfo {author} {\bibfnamefont {B.~A.}\ \bibnamefont {Dobrescu}},\ }\bibfield  {title} {\bibinfo {title} {{Massless gauge bosons other than the photon}},\ }\href {https://doi.org/10.1103/PHYSREVLETT.94.151802} {\bibfield  {journal} {\bibinfo  {journal} {Physical Review Letters}\ }\textbf {\bibinfo {volume} {94}},\ \bibinfo {pages} {151802} (\bibinfo {year} {2005})},\ \Eprint {https://arxiv.org/abs/hep-ph/0411004} {arXiv:hep-ph/0411004} \BibitemShut {NoStop}%
\bibitem [{\citenamefont {Dobrescu}\ and\ \citenamefont {Mocioiu}(2006)}]{Dobrescu2006Spin}%
  \BibitemOpen
  \bibfield  {author} {\bibinfo {author} {\bibfnamefont {B.~A.}\ \bibnamefont {Dobrescu}}\ and\ \bibinfo {author} {\bibfnamefont {I.}~\bibnamefont {Mocioiu}},\ }\bibfield  {title} {\bibinfo {title} {{Spin-dependent macroscopic forces from new particle exchange}},\ }\href {https://doi.org/10.1088/1126-6708/2006/11/005} {\bibfield  {journal} {\bibinfo  {journal} {Journal of High Energy Physics}\ }\textbf {\bibinfo {volume} {2006}},\ \bibinfo {pages} {005} (\bibinfo {year} {2006})},\ \Eprint {https://arxiv.org/abs/hep-ph/0605342} {arXiv:hep-ph/0605342} \BibitemShut {NoStop}%
\bibitem [{\citenamefont {Fabbrichesi}\ \emph {et~al.}(2020)\citenamefont {Fabbrichesi}, \citenamefont {Gabrielli},\ and\ \citenamefont {Lanfranchi}}]{Fabbrichesi2020TheDark}%
  \BibitemOpen
  \bibfield  {author} {\bibinfo {author} {\bibfnamefont {M.}~\bibnamefont {Fabbrichesi}}, \bibinfo {author} {\bibfnamefont {E.}~\bibnamefont {Gabrielli}},\ and\ \bibinfo {author} {\bibfnamefont {G.}~\bibnamefont {Lanfranchi}},\ }\bibfield  {title} {\bibinfo {title} {{The Dark Photon}},\ }\href {https://doi.org/10.1007/978-3-030-62519-1} {\bibfield  {journal} {\bibinfo  {journal} {SpringerBriefs in Physics 2020}\ ,\ \bibinfo {pages} {235}} (\bibinfo {year} {2020})},\ \Eprint {https://arxiv.org/abs/2005.01515} {arXiv:2005.01515} \BibitemShut {NoStop}%
\bibitem [{\citenamefont {An}\ \emph {et~al.}(2015)\citenamefont {An}, \citenamefont {Pospelov}, \citenamefont {Pradler},\ and\ \citenamefont {Ritz}}]{An2015Direct}%
  \BibitemOpen
  \bibfield  {author} {\bibinfo {author} {\bibfnamefont {H.}~\bibnamefont {An}}, \bibinfo {author} {\bibfnamefont {M.}~\bibnamefont {Pospelov}}, \bibinfo {author} {\bibfnamefont {J.}~\bibnamefont {Pradler}},\ and\ \bibinfo {author} {\bibfnamefont {A.}~\bibnamefont {Ritz}},\ }\bibfield  {title} {\bibinfo {title} {{Direct detection constraints on dark photon dark matter}},\ }\href {https://doi.org/10.1016/J.PHYSLETB.2015.06.018} {\bibfield  {journal} {\bibinfo  {journal} {Physics Letters B}\ }\textbf {\bibinfo {volume} {747}},\ \bibinfo {pages} {331} (\bibinfo {year} {2015})},\ \Eprint {https://arxiv.org/abs/1412.8378} {arXiv:1412.8378} \BibitemShut {NoStop}%
\bibitem [{\citenamefont {Kahn}\ \emph {et~al.}(2017)\citenamefont {Kahn}, \citenamefont {Krnjaic}, \citenamefont {Mishra-Sharma},\ and\ \citenamefont {Tait}}]{Kahn2017Light}%
  \BibitemOpen
  \bibfield  {author} {\bibinfo {author} {\bibfnamefont {Y.}~\bibnamefont {Kahn}}, \bibinfo {author} {\bibfnamefont {G.}~\bibnamefont {Krnjaic}}, \bibinfo {author} {\bibfnamefont {S.}~\bibnamefont {Mishra-Sharma}},\ and\ \bibinfo {author} {\bibfnamefont {T.~M.~P.}\ \bibnamefont {Tait}},\ }\bibfield  {title} {\bibinfo {title} {{Light weakly coupled axial forces: models, constraints, and projections}},\ }\href {https://doi.org/10.1007/JHEP05(2017)002} {\bibfield  {journal} {\bibinfo  {journal} {Journal of High Energy Physics}\ }\textbf {\bibinfo {volume} {2017}},\ \bibinfo {pages} {2} (\bibinfo {year} {2017})},\ \Eprint {https://arxiv.org/abs/1609.09072} {arXiv:1609.09072} \BibitemShut {NoStop}%
\bibitem [{\citenamefont {Dienes}\ and\ \citenamefont {Thomas}(2012)}]{Dienes2012Dynamical}%
  \BibitemOpen
  \bibfield  {author} {\bibinfo {author} {\bibfnamefont {K.~R.}\ \bibnamefont {Dienes}}\ and\ \bibinfo {author} {\bibfnamefont {B.}~\bibnamefont {Thomas}},\ }\bibfield  {title} {\bibinfo {title} {{Dynamical dark matter. I. Theoretical overview}},\ }\href {https://doi.org/10.1103/PHYSREVD.85.083523} {\bibfield  {journal} {\bibinfo  {journal} {Physical Review D}\ }\textbf {\bibinfo {volume} {85}},\ \bibinfo {pages} {083523} (\bibinfo {year} {2012})},\ \Eprint {https://arxiv.org/abs/1106.4546} {arXiv:1106.4546} \BibitemShut {NoStop}%
\bibitem [{\citenamefont {Heeck}(2014)}]{Heeck2014Unbroken}%
  \BibitemOpen
  \bibfield  {author} {\bibinfo {author} {\bibfnamefont {J.}~\bibnamefont {Heeck}},\ }\bibfield  {title} {\bibinfo {title} {{Unbroken B–L symmetry}},\ }\href {https://doi.org/10.1016/J.PHYSLETB.2014.10.067} {\bibfield  {journal} {\bibinfo  {journal} {Physics Letters B}\ }\textbf {\bibinfo {volume} {739}},\ \bibinfo {pages} {256} (\bibinfo {year} {2014})},\ \Eprint {https://arxiv.org/abs/1408.6845} {arXiv:1408.6845} \BibitemShut {NoStop}%
\bibitem [{\citenamefont {Peskin}\ and\ \citenamefont {Shroeder}(1995)}]{Peskin1995AnIntroduction}%
  \BibitemOpen
  \bibfield  {author} {\bibinfo {author} {\bibfnamefont {E.}~\bibnamefont {Peskin}, \bibfnamefont {Michael}}\ and\ \bibinfo {author} {\bibfnamefont {V.}~\bibnamefont {Shroeder}, \bibfnamefont {Daniel}},\ } {\emph {\bibinfo {title} {{An Introduction to Quantum Field Theory}}}}\ (\bibinfo  {publisher} {Perseus Books},\ \bibinfo {address} {Reading, MA},\ \bibinfo {year} {1995})\BibitemShut {NoStop}%
\bibitem [{\citenamefont {Clowe}\ \emph {et~al.}(2006)\citenamefont {Clowe}, \citenamefont {Brada{\v{c}}}, \citenamefont {Gonzalez}, \citenamefont {Markevitch}, \citenamefont {Randall}, \citenamefont {Jones},\ and\ \citenamefont {Zaritsky}}]{Clowe2006ADirect}%
  \BibitemOpen
  \bibfield  {author} {\bibinfo {author} {\bibfnamefont {D.}~\bibnamefont {Clowe}}, \bibinfo {author} {\bibfnamefont {M.}~\bibnamefont {Brada{\v{c}}}}, \bibinfo {author} {\bibfnamefont {A.~H.}\ \bibnamefont {Gonzalez}}, \bibinfo {author} {\bibfnamefont {M.}~\bibnamefont {Markevitch}}, \bibinfo {author} {\bibfnamefont {S.~W.}\ \bibnamefont {Randall}}, \bibinfo {author} {\bibfnamefont {C.}~\bibnamefont {Jones}},\ and\ \bibinfo {author} {\bibfnamefont {D.}~\bibnamefont {Zaritsky}},\ }\bibfield  {title} {\bibinfo {title} {{A Direct Empirical Proof of the Existence of Dark Matter*}},\ }\href {https://doi.org/10.1086/508162} {\bibfield  {journal} {\bibinfo  {journal} {The Astrophysical Journal}\ }\textbf {\bibinfo {volume} {648}},\ \bibinfo {pages} {L109} (\bibinfo {year} {2006})},\ \Eprint {https://arxiv.org/abs/astro-ph/0608407} {arXiv:astro-ph/0608407} \BibitemShut {NoStop}%
\bibitem [{\citenamefont {Roszkowski}\ \emph {et~al.}(2018)\citenamefont {Roszkowski}, \citenamefont {Sessolo},\ and\ \citenamefont {Trojanowski}}]{Roszkowski2018Wimp}%
  \BibitemOpen
  \bibfield  {author} {\bibinfo {author} {\bibfnamefont {L.}~\bibnamefont {Roszkowski}}, \bibinfo {author} {\bibfnamefont {E.~M.}\ \bibnamefont {Sessolo}},\ and\ \bibinfo {author} {\bibfnamefont {S.}~\bibnamefont {Trojanowski}},\ }\bibfield  {title} {\bibinfo {title} {{WIMP dark matter candidates and searches—current status and future prospects}},\ }\href {https://doi.org/10.1088/1361-6633/aab913} {\bibfield  {journal} {\bibinfo  {journal} {Reports on Progress in Physics}\ }\textbf {\bibinfo {volume} {81}},\ \bibinfo {pages} {66201} (\bibinfo {year} {2018})},\ \Eprint {https://arxiv.org/abs/1707.06277} {arXiv:1707.06277} \BibitemShut {NoStop}%
\bibitem [{\citenamefont {Navas}(2024)}]{Navas2024Review}%
  \BibitemOpen
  \bibfield  {author} {\bibinfo {author} {\bibfnamefont {S.}~\bibnamefont {Navas}},\ }\bibfield  {title} {\bibinfo {title} {{Review of Particle Physics}},\ }\href {https://doi.org/10.1103/PhysRevD.110.030001} {\bibfield  {journal} {\bibinfo  {journal} {Physical Review D}\ }\textbf {\bibinfo {volume} {110}},\ \bibinfo {pages} {030001} (\bibinfo {year} {2024})}\BibitemShut {NoStop}%
\bibitem [{\citenamefont {Kolb}\ and\ \citenamefont {Turner}(2018)}]{Kolb2018TheEarly}%
  \BibitemOpen
  \bibfield  {author} {\bibinfo {author} {\bibfnamefont {E.~W.}\ \bibnamefont {Kolb}}\ and\ \bibinfo {author} {\bibfnamefont {M.~S.}\ \bibnamefont {Turner}},\ }\href {https://doi.org/10.1201/9780429492860} {\emph {\bibinfo {title} {The Early Universe}}}\ (\bibinfo  {publisher} {CRC Press},\ \bibinfo {year} {2018})\ pp.\ \bibinfo {pages} {1--547}\BibitemShut {NoStop}%
\bibitem [{\citenamefont {Gondolo}\ and\ \citenamefont {Gelmini}(1991)}]{Gondolo1991Cosmic}%
  \BibitemOpen
  \bibfield  {author} {\bibinfo {author} {\bibfnamefont {P.}~\bibnamefont {Gondolo}}\ and\ \bibinfo {author} {\bibfnamefont {G.}~\bibnamefont {Gelmini}},\ }\bibfield  {title} {\bibinfo {title} {{Cosmic abundances of stable particles: Improved analysis}},\ }\href {https://doi.org/10.1016/0550-3213(91)90438-4} {\bibfield  {journal} {\bibinfo  {journal} {Nuclear Physics B}\ }\textbf {\bibinfo {volume} {360}},\ \bibinfo {pages} {145} (\bibinfo {year} {1991})}\BibitemShut {NoStop}%
\bibitem [{\citenamefont {Lee}\ and\ \citenamefont {Weinberg}(1977)}]{Lee1977Cosmological}%
  \BibitemOpen
  \bibfield  {author} {\bibinfo {author} {\bibfnamefont {B.~W.}\ \bibnamefont {Lee}}\ and\ \bibinfo {author} {\bibfnamefont {S.}~\bibnamefont {Weinberg}},\ }\bibfield  {title} {\bibinfo {title} {{Cosmological Lower Bound on Heavy-Neutrino Masses}},\ }\href {https://doi.org/10.1103/PhysRevLett.39.165} {\bibfield  {journal} {\bibinfo  {journal} {Physical Review Letters}\ }\textbf {\bibinfo {volume} {39}},\ \bibinfo {pages} {165} (\bibinfo {year} {1977})}\BibitemShut {NoStop}%
\bibitem [{\citenamefont {Steigman}\ \emph {et~al.}(2012)\citenamefont {Steigman}, \citenamefont {Dasgupta},\ and\ \citenamefont {Beacom}}]{Steigman2012Precise}%
  \BibitemOpen
  \bibfield  {author} {\bibinfo {author} {\bibfnamefont {G.}~\bibnamefont {Steigman}}, \bibinfo {author} {\bibfnamefont {B.}~\bibnamefont {Dasgupta}},\ and\ \bibinfo {author} {\bibfnamefont {J.~F.}\ \bibnamefont {Beacom}},\ }\bibfield  {title} {\bibinfo {title} {{Precise relic WIMP abundance and its impact on searches for dark matter annihilation}},\ }\href {https://doi.org/10.1103/PHYSREVD.86.023506} {\bibfield  {journal} {\bibinfo  {journal} {Physical Review D}\ }\textbf {\bibinfo {volume} {86}},\ \bibinfo {pages} {023506} (\bibinfo {year} {2012})},\ \Eprint {https://arxiv.org/abs/1204.3622} {arXiv:1204.3622} \BibitemShut {NoStop}%
\bibitem [{\citenamefont {Baer}\ \emph {et~al.}(2015)\citenamefont {Baer}, \citenamefont {Choi}, \citenamefont {Kim},\ and\ \citenamefont {Roszkowski}}]{Baer2015Dark}%
  \BibitemOpen
  \bibfield  {author} {\bibinfo {author} {\bibfnamefont {H.}~\bibnamefont {Baer}}, \bibinfo {author} {\bibfnamefont {K.~Y.}\ \bibnamefont {Choi}}, \bibinfo {author} {\bibfnamefont {J.~E.}\ \bibnamefont {Kim}},\ and\ \bibinfo {author} {\bibfnamefont {L.}~\bibnamefont {Roszkowski}},\ }\bibfield  {title} {\bibinfo {title} {{Dark matter production in the early Universe: Beyond the thermal WIMP paradigm}},\ }\href {https://doi.org/10.1016/j.physrep.2014.10.002} {\bibfield  {journal} {\bibinfo  {journal} {Physics Reports}\ }\textbf {\bibinfo {volume} {555}},\ \bibinfo {pages} {1} (\bibinfo {year} {2015})},\ \Eprint {https://arxiv.org/abs/1407.0017} {arXiv:1407.0017} \BibitemShut {NoStop}%
\bibitem [{\citenamefont {Heikinheimo}\ \emph {et~al.}(2017)\citenamefont {Heikinheimo}, \citenamefont {Tenkanen},\ and\ \citenamefont {Tuominen}}]{Heikinheimo2017Wimp}%
  \BibitemOpen
  \bibfield  {author} {\bibinfo {author} {\bibfnamefont {M.}~\bibnamefont {Heikinheimo}}, \bibinfo {author} {\bibfnamefont {T.}~\bibnamefont {Tenkanen}},\ and\ \bibinfo {author} {\bibfnamefont {K.}~\bibnamefont {Tuominen}},\ }\bibfield  {title} {\bibinfo {title} {{WIMP miracle of the second kind}},\ }\href {https://doi.org/10.1103/PHYSREVD.96.023001} {\bibfield  {journal} {\bibinfo  {journal} {Physical Review D}\ }\textbf {\bibinfo {volume} {96}},\ \bibinfo {pages} {023001} (\bibinfo {year} {2017})},\ \Eprint {https://arxiv.org/abs/1704.05359} {arXiv:1704.05359} \BibitemShut {NoStop}%
\bibitem [{\citenamefont {Hall}\ \emph {et~al.}(2010)\citenamefont {Hall}, \citenamefont {Jedamzik}, \citenamefont {March-Russell},\ and\ \citenamefont {West}}]{Hall2010Freeze}%
  \BibitemOpen
  \bibfield  {author} {\bibinfo {author} {\bibfnamefont {L.~J.}\ \bibnamefont {Hall}}, \bibinfo {author} {\bibfnamefont {K.}~\bibnamefont {Jedamzik}}, \bibinfo {author} {\bibfnamefont {J.}~\bibnamefont {March-Russell}},\ and\ \bibinfo {author} {\bibfnamefont {S.~M.}\ \bibnamefont {West}},\ }\bibfield  {title} {\bibinfo {title} {{Freeze-in production of FIMP dark matter}},\ }\href {https://doi.org/10.1007/JHEP03(2010)080} {\bibfield  {journal} {\bibinfo  {journal} {Journal of High Energy Physics}\ }\textbf {\bibinfo {volume} {2010}},\ \bibinfo {pages} {1} (\bibinfo {year} {2010})},\ \Eprint {https://arxiv.org/abs/0911.1120} {arXiv:0911.1120} \BibitemShut {NoStop}%
\bibitem [{\citenamefont {Bernal}\ \emph {et~al.}(2017)\citenamefont {Bernal}, \citenamefont {Heikinheimo}, \citenamefont {Tenkanen}, \citenamefont {Tuominen},\ and\ \citenamefont {Vaskonen}}]{Bernal2017TheDawn}%
  \BibitemOpen
  \bibfield  {author} {\bibinfo {author} {\bibfnamefont {N.}~\bibnamefont {Bernal}}, \bibinfo {author} {\bibfnamefont {M.}~\bibnamefont {Heikinheimo}}, \bibinfo {author} {\bibfnamefont {T.}~\bibnamefont {Tenkanen}}, \bibinfo {author} {\bibfnamefont {K.}~\bibnamefont {Tuominen}},\ and\ \bibinfo {author} {\bibfnamefont {V.}~\bibnamefont {Vaskonen}},\ }\bibfield  {title} {\bibinfo {title} {{The dawn of FIMP Dark Matter: A review of models and constraints}},\ }\bibfield  {journal} {\bibinfo  {journal} {International Journal of Modern Physics A}\ }\textbf {\bibinfo {volume} {32}},\ \href {https://doi.org/10.1142/S0217751X1730023X} {10.1142/S0217751X1730023X} (\bibinfo {year} {2017}),\ \Eprint {https://arxiv.org/abs/1706.07442} {arXiv:1706.07442} \BibitemShut {NoStop}%
\bibitem [{\citenamefont {B{\'{e}}langer}\ \emph {et~al.}(2020)\citenamefont {B{\'{e}}langer}, \citenamefont {Delaunay}, \citenamefont {Pukhov},\ and\ \citenamefont {Zaldivar}}]{Belanger2020Dark}%
  \BibitemOpen
  \bibfield  {author} {\bibinfo {author} {\bibfnamefont {G.}~\bibnamefont {B{\'{e}}langer}}, \bibinfo {author} {\bibfnamefont {C.}~\bibnamefont {Delaunay}}, \bibinfo {author} {\bibfnamefont {A.}~\bibnamefont {Pukhov}},\ and\ \bibinfo {author} {\bibfnamefont {B.}~\bibnamefont {Zaldivar}},\ }\bibfield  {title} {\bibinfo {title} {{Dark matter abundance from the sequential freeze-in mechanism}},\ }\href {https://doi.org/10.1103/PHYSREVD.102.035017} {\bibfield  {journal} {\bibinfo  {journal} {Physical Review D}\ }\textbf {\bibinfo {volume} {102}},\ \bibinfo {pages} {035017} (\bibinfo {year} {2020})},\ \Eprint {https://arxiv.org/abs/2005.06294} {arXiv:2005.06294} \BibitemShut {NoStop}%
\bibitem [{\citenamefont {Du}\ \emph {et~al.}(2022)\citenamefont {Du}, \citenamefont {Huang}, \citenamefont {Li}, \citenamefont {Li},\ and\ \citenamefont {Yu}}]{Du2022Revisiting}%
  \BibitemOpen
  \bibfield  {author} {\bibinfo {author} {\bibfnamefont {Y.}~\bibnamefont {Du}}, \bibinfo {author} {\bibfnamefont {F.}~\bibnamefont {Huang}}, \bibinfo {author} {\bibfnamefont {H.~L.}\ \bibnamefont {Li}}, \bibinfo {author} {\bibfnamefont {Y.~Z.}\ \bibnamefont {Li}},\ and\ \bibinfo {author} {\bibfnamefont {J.~H.}\ \bibnamefont {Yu}},\ }\bibfield  {title} {\bibinfo {title} {{Revisiting dark matter freeze-in and freeze-out through phase-space distribution}},\ }\href {https://doi.org/10.1088/1475-7516/2022/04/012} {\bibfield  {journal} {\bibinfo  {journal} {JCAP}\ }\textbf {\bibinfo {volume} {2022}}\bibfield  {number} {\bibinfo  {number} { (04)},\ \bibinfo {pages} {012}},\ }\Eprint {https://arxiv.org/abs/2111.01267} {arXiv:2111.01267} \BibitemShut {NoStop}%
\bibitem [{\citenamefont {Laine}\ and\ \citenamefont {Schr{\"{o}}der}(2006)}]{Laine2006Quark}%
  \BibitemOpen
  \bibfield  {author} {\bibinfo {author} {\bibfnamefont {M.}~\bibnamefont {Laine}}\ and\ \bibinfo {author} {\bibfnamefont {Y.}~\bibnamefont {Schr{\"{o}}der}},\ }\bibfield  {title} {\bibinfo {title} {{Quark mass thresholds in QCD thermodynamics}},\ }\href {https://doi.org/10.1103/PHYSREVD.73.085009} {\bibfield  {journal} {\bibinfo  {journal} {Physical Review D}\ }\textbf {\bibinfo {volume} {73}},\ \bibinfo {pages} {085009} (\bibinfo {year} {2006})},\ \Eprint {https://arxiv.org/abs/hep-ph/0603048} {arXiv:hep-ph/0603048} \BibitemShut {NoStop}%
\bibitem [{\citenamefont {Frumkin}\ \emph {et~al.}(2023)\citenamefont {Frumkin}, \citenamefont {Hochberg}, \citenamefont {Kuflik},\ and\ \citenamefont {Murayama}}]{Frumkin2023Thermal}%
  \BibitemOpen
  \bibfield  {author} {\bibinfo {author} {\bibfnamefont {R.}~\bibnamefont {Frumkin}}, \bibinfo {author} {\bibfnamefont {Y.}~\bibnamefont {Hochberg}}, \bibinfo {author} {\bibfnamefont {E.}~\bibnamefont {Kuflik}},\ and\ \bibinfo {author} {\bibfnamefont {H.}~\bibnamefont {Murayama}},\ }\bibfield  {title} {\bibinfo {title} {{Thermal Dark Matter from Freeze-Out of Inverse Decays}},\ }\href {https://doi.org/10.1103/PHYSREVLETT.130.121001} {\bibfield  {journal} {\bibinfo  {journal} {Physical Review Letters}\ }\textbf {\bibinfo {volume} {130}},\ \bibinfo {pages} {121001} (\bibinfo {year} {2023})},\ \Eprint {https://arxiv.org/abs/2111.14857} {arXiv:2111.14857} \BibitemShut {NoStop}%
\bibitem [{\citenamefont {Cahn}(1989)}]{Cahn1989TheHiggs}%
  \BibitemOpen
  \bibfield  {author} {\bibinfo {author} {\bibfnamefont {R.~N.}\ \bibnamefont {Cahn}},\ }\bibfield  {title} {\bibinfo {title} {{The Higgs boson}},\ }\href {https://doi.org/10.1088/0034-4885/52/4/001} {\bibfield  {journal} {\bibinfo  {journal} {Reports on Progress in Physics}\ }\textbf {\bibinfo {volume} {52}},\ \bibinfo {pages} {389} (\bibinfo {year} {1989})}\BibitemShut {NoStop}%
\bibitem [{\citenamefont {Grau}\ \emph {et~al.}(1990)\citenamefont {Grau}, \citenamefont {Pancheri},\ and\ \citenamefont {Philips}}]{Grau1990Contributions}%
  \BibitemOpen
  \bibfield  {author} {\bibinfo {author} {\bibfnamefont {A.}~\bibnamefont {Grau}}, \bibinfo {author} {\bibfnamefont {G.}~\bibnamefont {Pancheri}},\ and\ \bibinfo {author} {\bibfnamefont {R.~J.}\ \bibnamefont {Philips}},\ }\bibfield  {title} {\bibinfo {title} {{Contributions of off-shell top quarks to decay processes}},\ }\href {https://doi.org/10.1016/0370-2693(90)90939-4} {\bibfield  {journal} {\bibinfo  {journal} {Physics Letters B}\ }\textbf {\bibinfo {volume} {251}},\ \bibinfo {pages} {293} (\bibinfo {year} {1990})}\BibitemShut {NoStop}%
\bibitem [{\citenamefont {Rom{\~{a}}o}\ and\ \citenamefont {Andringa}(1999)}]{Romao1999Vector}%
  \BibitemOpen
  \bibfield  {author} {\bibinfo {author} {\bibfnamefont {J.~C.}\ \bibnamefont {Rom{\~{a}}o}}\ and\ \bibinfo {author} {\bibfnamefont {S.}~\bibnamefont {Andringa}},\ }\bibfield  {title} {\bibinfo {title} {{Vector boson decays of the Higgs boson}},\ }\href {https://doi.org/10.1007/S100529801038} {\bibfield  {journal} {\bibinfo  {journal} {European Physical Journal C}\ }\textbf {\bibinfo {volume} {7}},\ \bibinfo {pages} {631} (\bibinfo {year} {1999})},\ \Eprint {https://arxiv.org/abs/hep-ph/9807536} {arXiv:hep-ph/9807536} \BibitemShut {NoStop}%
\bibitem [{\citenamefont {Djouadi}(2008)}]{Djouadi2008TheAnatomy}%
  \BibitemOpen
  \bibfield  {author} {\bibinfo {author} {\bibfnamefont {A.}~\bibnamefont {Djouadi}},\ }\bibfield  {title} {\bibinfo {title} {{The anatomy of electroweak symmetry breaking: Tome I: The Higgs boson in the Standard Model}},\ }\href {https://doi.org/10.1016/J.PHYSREP.2007.10.004} {\bibfield  {journal} {\bibinfo  {journal} {Physics Reports}\ }\textbf {\bibinfo {volume} {457}},\ \bibinfo {pages} {1} (\bibinfo {year} {2008})},\ \Eprint {https://arxiv.org/abs/hep-ph/0503172} {arXiv:hep-ph/0503172} \BibitemShut {NoStop}%
\bibitem [{\citenamefont {Choi}\ \emph {et~al.}(2021)\citenamefont {Choi}, \citenamefont {Lee},\ and\ \citenamefont {Park}}]{Choi2021Decays}%
  \BibitemOpen
  \bibfield  {author} {\bibinfo {author} {\bibfnamefont {S.~Y.}\ \bibnamefont {Choi}}, \bibinfo {author} {\bibfnamefont {J.~S.}\ \bibnamefont {Lee}},\ and\ \bibinfo {author} {\bibfnamefont {J.}~\bibnamefont {Park}},\ }\bibfield  {title} {\bibinfo {title} {{Decays of Higgs bosons in the Standard Model and beyond}},\ }\href {https://doi.org/10.1016/j.ppnp.2021.103880} {\bibfield  {journal} {\bibinfo  {journal} {PPNP}\ }\textbf {\bibinfo {volume} {120}},\ \bibinfo {pages} {103880} (\bibinfo {year} {2021})},\ \Eprint {https://arxiv.org/abs/2101.12435} {arXiv:2101.12435} \BibitemShut {NoStop}%
\bibitem [{\citenamefont {Randall}\ \emph {et~al.}(2008)\citenamefont {Randall}, \citenamefont {Markevitch}, \citenamefont {Clowe}, \citenamefont {Gonzalez},\ and\ \citenamefont {Brada{\v{c}}}}]{Randall2008Constraints}%
  \BibitemOpen
  \bibfield  {author} {\bibinfo {author} {\bibfnamefont {S.~W.}\ \bibnamefont {Randall}}, \bibinfo {author} {\bibfnamefont {M.}~\bibnamefont {Markevitch}}, \bibinfo {author} {\bibfnamefont {D.}~\bibnamefont {Clowe}}, \bibinfo {author} {\bibfnamefont {A.~H.}\ \bibnamefont {Gonzalez}},\ and\ \bibinfo {author} {\bibfnamefont {M.}~\bibnamefont {Brada{\v{c}}}},\ }\bibfield  {title} {\bibinfo {title} {{Constraints on the Self‐Interaction Cross Section of Dark Matter from Numerical Simulations of the Merging Galaxy Cluster 1E 0657-56}},\ }\href {https://doi.org/10.1086/587859} {\bibfield  {journal} {\bibinfo  {journal} {The Astrophysical Journal}\ }\textbf {\bibinfo {volume} {679}},\ \bibinfo {pages} {1173} (\bibinfo {year} {2008})},\ \Eprint {https://arxiv.org/abs/0704.0261} {arXiv:0704.0261} \BibitemShut {NoStop}%
\bibitem [{\citenamefont {Tulin}\ and\ \citenamefont {Yu}(2018)}]{Tulin2018Dark}%
  \BibitemOpen
  \bibfield  {author} {\bibinfo {author} {\bibfnamefont {S.}~\bibnamefont {Tulin}}\ and\ \bibinfo {author} {\bibfnamefont {H.~B.}\ \bibnamefont {Yu}},\ }\bibfield  {title} {\bibinfo {title} {{Dark matter self-interactions and small scale structure}},\ }\href {https://doi.org/10.1016/j.physrep.2017.11.004} {\bibfield  {journal} {\bibinfo  {journal} {Physics Reports}\ }\textbf {\bibinfo {volume} {730}},\ \bibinfo {pages} {1} (\bibinfo {year} {2018})},\ \Eprint {https://arxiv.org/abs/1705.02358} {arXiv:1705.02358} \BibitemShut {NoStop}%
\bibitem [{\citenamefont {Harvey}\ \emph {et~al.}(2015)\citenamefont {Harvey}, \citenamefont {Massey}, \citenamefont {Kitching}, \citenamefont {Taylor},\ and\ \citenamefont {Tittley}}]{Harvey2015TheNongravitational}%
  \BibitemOpen
  \bibfield  {author} {\bibinfo {author} {\bibfnamefont {D.}~\bibnamefont {Harvey}}, \bibinfo {author} {\bibfnamefont {R.}~\bibnamefont {Massey}}, \bibinfo {author} {\bibfnamefont {T.}~\bibnamefont {Kitching}}, \bibinfo {author} {\bibfnamefont {A.}~\bibnamefont {Taylor}},\ and\ \bibinfo {author} {\bibfnamefont {E.}~\bibnamefont {Tittley}},\ }\bibfield  {title} {\bibinfo {title} {{The nongravitational interactions of dark matter in colliding galaxy clusters}},\ }\href {https://doi.org/10.1126/SCIENCE.1261381} {\bibfield  {journal} {\bibinfo  {journal} {Science}\ }\textbf {\bibinfo {volume} {347}},\ \bibinfo {pages} {1462} (\bibinfo {year} {2015})},\ \Eprint {https://arxiv.org/abs/1503.07675} {arXiv:1503.07675} \BibitemShut {NoStop}%

\end{thebibliography}

\end{document}